\documentclass{llncs}

\usepackage[utf8]{inputenc}
\usepackage{amsmath,amssymb,amsfonts}
\usepackage{algorithm}
\usepackage{algpseudocode}

\usepackage{graphicx}
\usepackage{textcomp}
\usepackage{xcolor}
\usepackage{colortbl}

\usepackage{siunitx}
\usepackage{microtype}                 
\usepackage{tabu}                      
\usepackage{booktabs}    
\usepackage{todonotes}

\usepackage{hyperref}
\usepackage{ dsfont }
\usepackage{xspace}
\usepackage{multirow}
\usepackage{subcaption}
\usepackage[labelfont=bf,labelsep=period]{caption}
\usepackage{etoolbox}
\usepackage[bottom]{footmisc}
\usepackage{bm}
\usepackage{enumitem}
\usepackage{rotating}
\usepackage{makecell}
\usepackage{stmaryrd}
\usepackage{hyperref}
\usepackage[sort]{cite}

\newcommand{\custRef}[2]{\hyperref[#1]{#2 \ref*{#1}}}

\newcommand{\ie}{\textit{i.e.},\xspace}

\newcommand{\eg}{\textit{e.g.},\xspace}

\newcommand{\etal}[1]{\textit{et al.}~\cite{#1}\xspace}

\newcommand{\CardV}{$|$V$|$\xspace}
\newcommand{\CardE}{$|$E$|$\xspace}
\newcommand{\pair}{(v_i, v_j)}

\newcommand{\sgdgd}{\textit{S\_GD\textsuperscript{2}}\xspace}

\newcommand{\modelname}{FORBID\xspace}
\newcommand{\modelnamep}{\modelname'\xspace}

\newcommand{\oonni}{\textit{\hyperref[met:oonni]{oo\_nni}}\xspace}
\newcommand{\spcha}{\textit{\hyperref[met:spcha]{sp\_ch\_a}}\xspace}
\newcommand{\gsbbiar}{\textit{\hyperref[met:gsbbiar]{gs\_bb\_iar}}\xspace}
\newcommand{\dmimse}{\textit{\hyperref[met:dmimse]{nm\_dm\_imse}}\xspace}
\newcommand{\elrsd}{\textit{\hyperref[met:elrsd]{el\_rsdd}}\xspace}

\makeatletter
\newcommand*{\algrule}[1][\algorithmicindent]{\makebox[#1][l]{\hspace*{.5em}\vrule height .75\baselineskip depth .25\baselineskip}}%

\newcount\ALG@printindent@tempcnta
\def\ALG@printindent{%
    \ifnum \theALG@nested>0
        \ifx\ALG@text\ALG@x@notext
            \addvspace{-3pt}
        \else
            \unskip
            \ALG@printindent@tempcnta=1
            \loop
                \algrule[\csname ALG@ind@\the\ALG@printindent@tempcnta\endcsname]%
                \advance \ALG@printindent@tempcnta 1
            \ifnum \ALG@printindent@tempcnta<\numexpr\theALG@nested+1\relax
            \repeat
        \fi
    \fi
    }%
\patchcmd{\ALG@doentity}{\noindent\hskip\ALG@tlm}{\ALG@printindent}{}{\errmessage{failed to patch}}
\makeatother

\algdef{SE}[VARIABLES]{Variables}{EndVariables}
   {\algorithmicvariables}
   {\algorithmicend\ \algorithmicvariables}
\algnewcommand{\algorithmicvariables}{\textbf{Variables}}

\algdef{SE}[METHODS]{Methods}{EndMethods}
   {\algorithmicmethods}
   {\algorithmicend\ \algorithmicmethods}
\algnewcommand{\algorithmicmethods}{\textbf{Methods}}

\algnewcommand{\algorithmicand}{\textbf{and}\xspace}
\algnewcommand{\algorithmicor}{\textbf{or}\xspace}
\algnewcommand{\OR}{\algorithmicor}
\algnewcommand{\AND}{\algorithmicand}

\title{\modelname: Fast Overlap Removal By stochastic gradIent Descent for Graph Drawing}
\titlerunning{\modelname}  
\author{Loann~Giovannangeli\orcidID{0000-0002-9395-6495} \and
Frederic~Lalanne\orcidID{0000-0001-9108-0955} \and
Romain~Giot\orcidID{0000-0002-0638-7504} \and
Romain~Bourqui\orcidID{0000-0002-1847-2589}
}
\authorrunning{Loann~Giovannangeli et al.} 

\institute{LaBRI, UMR CNRS 5800, University Bordeaux, 33405 Talence, France\\
\email{\{firstname\}.\{lastname\}@u-bordeaux.fr}
}

\begin{document}

\maketitle              

\begin{abstract}
While many graph drawing algorithms consider nodes as points, graph visualization tools often represent them as shapes. These shapes support the display of information such as labels or encode various data with size or color. However, they can create overlaps between nodes which hinder the exploration process by hiding parts of the information. 
It is therefore of utmost importance to remove these overlaps to improve graph visualization readability. 
If not handled by the layout process, Overlap Removal (OR) algorithms have been proposed as layout post-processing. As graph layouts usually convey information about their topology, it is important that OR algorithms preserve them as much as possible. We propose a novel algorithm that models OR as a joint stress and scaling optimization problem, and leverages efficient stochastic gradient descent. This approach is compared with state-of-the-art algorithms, and several quality metrics demonstrate its efficiency to quickly remove overlaps while retaining the initial layout structures.
\keywords{Layout adjustment \and Overlap removal \and Stress optimization \and Stochastic gradient descent}
\end{abstract}

\section{Introduction}
\label{sec:introduction}
Most dimension reduction algorithms consider data as points (\eg Multi Dimensional Scaling~\cite{van2008tsne,tipping1999pca}, Graph Layout~\cite{zheng2018sgdgd,hachul2004fm3,kamada1989kawai}). However, most visualization tools represent them by shapes with an area, whether to encode additional data within the screen representation of the nodes, or because it is simply more visually pleasing for end-users. Rendering such data that was laid out as points with shapes creates \textit{overlaps} that can severely hinder the representation readability by hiding information. If not handled directly inside the dimension reduction algorithm (\eg \cite{kamps1995constraint,harel2002drawing}), it is then the responsibility of a post-processing \textit{Overlap Removal} (OR) algorithm to remove these overlaps.

In this paper, we consider the OR problem in graph layouts context, meaning that our laid out data points are nodes positioned by a graph drawing algorithm. These nodes are represented as rectangles with a position and size in two dimensions. Other shapes can be considered as well, as long as it is possible to check for overlaps and measure a distance between them. We consider that a pair of nodes overlaps if the intersection of their shapes representations is not null. 

An OR algorithm takes a set of nodes positions and sizes as input and move these nodes to remove \textit{all} overlaps while optimizing two main criteria: \textit{compactness} and \textit{initial layout preservation}. 
An algorithm that uniformly upscales the initial layout until there is no overlap perfectly works (\eg uniform Scaling~\cite{chen2020survey}), but its result is not satisfactory as it produces very sparse layouts where the nodes visible areas are significantly reduced. Hence, a good OR algorithm should be able to optimize \textit{compactness} by preserving the scale of the initial layout, using its empty spaces to move nodes apart. 
Initial layout preservation is also of utmost importance in a graph drawing context since the initial graph layout is often computed to emphasize the graph structure. It is then imperative to preserve the mental map a user has of the initial layout. 
In addition, graph representations also include visualization of their edges. In that regard, optimizing compactness only is not suited to the graph context since it ends up hiding the graph edges.

The main contribution of this paper is \modelname\footnote{FORBID implementation: \url{https://github.com/LoannGio/FORBID}}: Fast Overlap Removal By stochastic gradIent Descent, a novel OR algorithm dedicated to graph drawings that produces overlap-free layouts balancing compactness and initial layout preservation. To the best of our knowledge, it is the first method to explicitly optimize the conjunction of these two aspects. It models the problem as a \textit{stress} function: each pair of nodes in the overlap-free layout should be put at an \textit{ideal distance} such that (i) there is no longer overlaps in the layouts and (ii) the distances between all the nodes are preserved. This stress is then optimized with an efficient state-of-the-art stochastic gradient descent algorithm~\cite{zheng2018sgdgd}. Leveraging Chen~\etal{chen2019surveyShort,chen2020survey} evaluation protocol, \modelname is compared with major state-of-the-art OR algorithms on a set of quality metrics specifically selected for this purpose. It demonstrates great capabilities to preserve initial layouts while retaining a decent level of compactness.

The remainder of this paper is organized as follows.
\autoref{sec:related_works} presents related works principally centered around the description of OR algorithms and their evaluation as proposed in~\cite{chen2019surveyShort,chen2020survey}. \autoref{sec:framework} describes \modelname algorithm, while \autoref{sec:benchmark} reports its evaluation. Finally, \autoref{sec:discussion} discusses visual examples of overlap-free layouts from several OR algorithms as well as \modelname convergence. 

\textbf{Notations:} Let $G=(V, E)$ be a graph with $V =\{v_1, v_2, ..., v_{N}\}$ its set of $N=|V|$ nodes and $E \subseteq V \times V$ its set of edges. A graph layout is defined as a tensor $X \in \mathds{R}^{N \times 2}$ where $X_i$ is the node $v_i$ projection in 2D. A node $v_i$ is defined by a rectangle of width and height $(w_i, h_i)$ centered in $X_i$. Two nodes overlap each other if the intersection between their rectangles is not null. For convenience, we define the set of overlapping pairs of nodes in a graph by $O \subseteq V \times V$. A corresponding overlap-free layout is defined as $X' \in \mathds{R}^{N \times 2}$. The euclidean distance bewteen two nodes $v_i$ and $v_j$ is noted $||X_i - X_j||$.

\section{Related Works}
\label{sec:related_works}
This section presents major prior works in the Overlap Removal (OR) field. 
As described in Chen \textit{et al.} survey~\cite{chen2019surveyShort,chen2020survey}, several efficient OR algorithms exist, many of which (\eg \cite{huang2007fta,hayashi1998pfsprime,misue1995pfs}) rely on \textit{scan line}~\cite{dwyer2005vpsc} to detect overlap presency in $\mathcal{O}(N \log N)$ and find all overlaps in $\mathcal{O}(|O| N (\log N+|O|))$ which \textit{can} be faster than the pairwise search in $\mathcal{O}(N^2)$. 
These OR algorithms focus on different contexts (\eg graph or generic 2D representation) and optimize different criteria (\eg compactness, layout preservation). 
PFS~\cite{misue1995pfs}, PFS'~\cite{hayashi1998pfsprime}, FTA~\cite{huang2007fta}, RWordle-L~\cite{strobelt2012rwordle} and uniform scaling~\cite{chen2019surveyShort,chen2020survey} rely on the \textit{scan line} algorithm to remove overlaps. PFS, PFS' and FTA are made of two passes handling horizontal and vertical movements separately, while \mbox{RWordle-L} moves nodes on both axes at the same time. In the end, they all have a quadratic complexity according to how nodes movements are computed.
PRISM~\cite{gansner2010prism} models OR as a stress optimization problem in a \textit{proximity graph} (\ie Delaunay triangulation of the initial layout) of a layout and runs in $\mathcal{O}(t(mkN + N \log N))$ where $m, k$ are optimization hyper-parameters and $t$ depends on the number of overlaps. GTREE~\cite{nachmanson2016gtree} leverages PRISM proximity graph to remove overlaps, but constructs a minimum spanning tree upon it to reduce the number of forces to compute. They both propose a good level of initial layout preservation. As \modelname idea is close to that of PRISM in some way, it will be further discussed in \autoref{sec:framework}.
VPSC~\cite{dwyer2005vpsc,dwyer2006vpscerratum} models OR as a set of constraints to relax but tends to highly deform the initial layout. Its complexity is $\mathcal{O}(CN \log C)$ where $C$ is the number of constraints in $\mathcal{O}(N)$ to relax; leading to a final complexity in $\mathcal{O}(N^2 \log N)$.
Finally, Diamond~\cite{meulemans2019diamond} is another constraint programming-based OR algorithm in $\mathcal{O}(N^2)$ that optimizes orthogonal order preservation. Its originality is to propose to temporarily rotate nodes by $45^{\circ}$, representing them as \textit{diamonds} to facilitate the constraints relaxations.



\section{\modelname Algorithm}
\label{sec:framework}
\begin{figure}
    \centering
    \includegraphics[width=\linewidth]{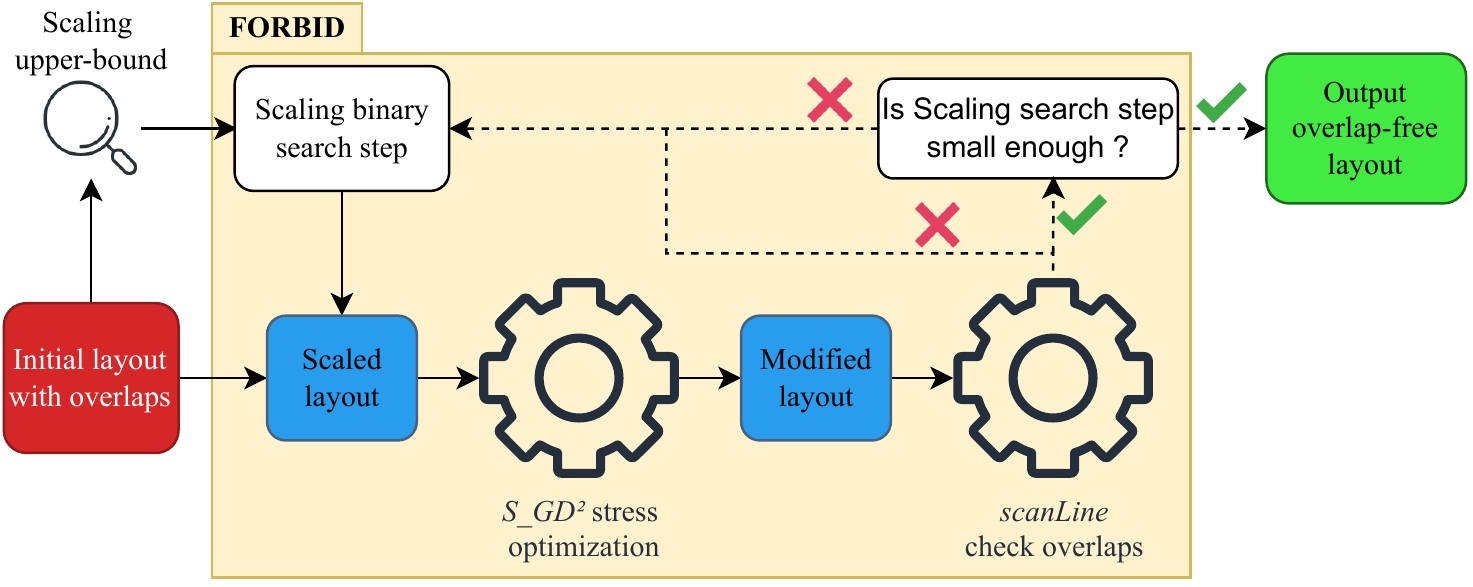}
    \caption{Simplified schema of \modelname algorithm representing how the stress and upscaling criteria are organized together. Gears represent the algorithms \modelname relies on (\ie \sgdgd~\cite{zheng2018sgdgd} and \textit{scan line}~\cite{dwyer2005vpsc}), and colored boxes represent layouts while white boxes represent the search for the optimal scaling.}
    \label{fig:method_schema}
\end{figure}

This section presents \modelname Overlap Removal (OR) algorithm. It is based on finding an optimal (\ie smallest) upscaling ratio while minimizing a stress function that models an overlap-free layout that preserves the initial one. The optimal scaling ratio is found with binary search, while the stress function is optimized with the \sgdgd algorithm~\cite{zheng2018sgdgd} that simulates stochastic gradient descent. An overview of the algorithm components is presented in \autoref{fig:method_schema} and its complexity is in $\mathcal{O}(s(N^2 + N \log N))$ where $s$ is defined later in \autoref{sec:scaling}.

\subsection{Stress Modelization for Overlap Removal}
\label{sec:stress_model}

\subsubsection{Preliminaries.}

Traditional graph layout algorithms (\eg~\cite{kamada1989kawai,brandes2008stress,ortmann2016stress,zheng2018sgdgd}) often optimize a \textit{stress} function that has been shown to lead to meaningful layouts and is defined as:

\begin{equation}
\label{eq:stress}
    \sigma(X) = \sum\limits_{i,j \in V} W_{ij}(||X_i - X_j|| - \delta_{ij})^2
\end{equation}
where $\delta_{ij}$ is an \textit{ideal distance} that the projected layout should preserve, and $W_{ij}$ is a weight factor usually set to $\delta_{ij}^{-2}$. In the graph layout context, $\delta_{ij}$ is set to the graph theoretical distances so that the projected representation of the graph enables end-users to apprehend the graph structure.

As proposed by Gansner and Hu~\cite{gansner2010prism}, stress is also a good criterion for OR problems. In fact, optimizing stress comes down to fit a distribution of distances in a low dimensional space to that of a higher space, considered ideal but of too high dimensions to be represented. The adaptation to an overlap removal context simply changes the notion of \textit{ideal distances}. Rather than fitting the projected positions to match graph theoretical distances, the ideal distances are two-folded here. For a pair of nodes $p_{ij} = \pair$, with $p_{ij} \notin O$, the ideal distance $\delta_{ij}$ is set to their distance $||X_i - X_j||$ in the original layout. On the other hand, if $p_{ij} \in O$, $\delta_{ij}$ is set to a distance such that $v_i$ and $v_j$ do not overlap anymore; that distance depending on the nodes \textit{shapes} and some design choices. This two-folded definition enables the optimization of both the preservation of the original layout and the overlapped nodes movement at the same time. 

PRISM~\cite{gansner2010prism} is an example of OR algorithm that optimizes stress. It constructs a \textit{proximity graph} (Delaunay triangulation of the initial layout) and optimizes stress alongside its edges. Considering nodes are represented as rectangles, they define the ideal distance as $\delta_{ij} = s_{ij}||X_i - X_j||$ where $s_{ij}$ is an expansion factor of the edge $\pair$ computed so that both nodes would be side by side. PRISM main limitations comes from the use of the \textit{proximity graph} that does not capture all the overlaps and only enables the preservation of distances between close nodes in the initial layout, ignoring longer distances preservation.

\subsubsection{\modelname Stress Modelization.}
For every pair of overlapped nodes, $\delta_{ij}$ is set to the distance between $v_i$ and $v_j$ centers if they were tangent in their corner. It allows one of them to be placed on the circle centered on the other and which radius is the minimum so that the nodes do not overlap anymore regardless of their relative position. 
This distance ensures that two nodes do not overlap anymore and favors convergence by adding some margin space them, which is necessary for two reasons. First, since stress is optimized by stochastic gradient descent in \modelname (see \autoref{sec:sgd2_optim}), the ideal distances will most likely get approximated, but never perfectly matched. Second, the PRISM definition of ideal distance means there is only one correct placement for every pair of overlapping nodes that satisfies the stress. However, as opposed to PRISM, \modelname optimizes the distances between all $N^2$ pairs of nodes, making the number of constraints to relax much higher. This margin enables \modelname to converge faster toward a solution. Formally, the stress is defined as \autoref{eq:stress} with $\delta_{ij}$ and $W_{ij}$ set to:

\begin{equation}
    \label{eq:delta_ij}
    \delta_{ij} =
    \begin{cases}
        \sqrt{\left(\frac{w_i + w_j}{2}\right)^2 +\left(\frac{h_i + h_j}{2}\right)^2}, \textrm{if }(v_i, v_j) \in O \\[10pt]
        ||X_i - X_j||, \textrm{if }(v_i, v_j) \notin O
    \end{cases}  
\end{equation}

\begin{equation}
    \label{eq:w_ij}
    W_{ij} =
    \begin{cases}
        \delta_{ij}^{k*\alpha}, \textrm{if } (v_i, v_j) \in O \\[10pt]
        \delta_{ij}^{\alpha}, \textrm{if } (v_i, v_j) \notin O
    \end{cases}  
\end{equation}
where $\alpha$ is generally set to $-2$ and $k \in \mathbb{R}$ is an overlap-related factor to tailor the algorithm behavior to a desired initial layout preservation. The smaller the weight is given to overlapping pairs of nodes, the more the initial layout will be preserved at the cost of slower convergence or higher scaling. 


\subsection{\modelname Stress Optimization}
\label{sec:sgd2_optim}

\modelname optimizes stress by stochastic gradient descent by leveraging \sgdgd~\cite{zheng2018sgdgd} algorithm. \sgdgd models stress as a set of constraints that are relaxed by individually moving pairs of nodes. The process is based on constrained graph layout~\cite{dwyer2009scalable,bostock2011d3}, and is optimized by considering the constraints individually to efficiently model clothes movements~\cite{jakobsen2001advanced}. By individually moving pairs of nodes to optimize the stress, the algorithm can create new overlaps. And since overlapping and non-overlapping pairs of nodes have different notions of ideal distances, $\delta_{ij}$ and $W_{ij}$ are updated at each optimization iteration (everytime all pairs of nodes have been moved once) so that, at any time, the algorithm optimizes distances according to the current state of the layout.

\sgdgd optimization convergence is based on an annealing step size schedule that mimicks a stochastic gradient descent. It computes $\mathcal{O}(N^2)$ movements, but in practice it converges in very few iterations, making it competitive with other algorithms (see \autoref{sec:exec_time}). To make it even faster, we also stop the gradient descent if, at any iteration, the sum of all nodes movements in the current iteration is null. This explains why every execution of \sgdgd is not necessarily of the same length (see \autoref{sec:convergence}).

\subsection{Scaling to Ensure Convergence}
\label{sec:scaling}

Here, we define the \textit{bounding box} as the minimum rectangle in which the initial layout fits.
Removing all overlaps in a layout is sometimes not feasible without deforming its bounding box, \eg when the sum of the nodes areas exceeds the layout bounding box area. Two strategies can overcome this: (i) to allow the optimization algorithm to distort the bounding shape, or (ii) to scale up the drawing to create empty spaces that can be used to move nodes. These two choices boil down to considering that either ``more space is needed'' or ``nodes must be smaller'' to provide an overlap-free layout. The first strategy has the benefit of limiting the bounding box upscaling, but can result in strongly distorted layouts that makes it difficult to recognize the original graph structure. To guarantee that \modelname finds a solution, it uses the second method and searches for the optimal upscale ratio so that there is enough space to remove overlaps without deforming its aspect. The optimal upscale ratio $s_{max}$ is found by binary search between $1$ and the minimum scaling ratio for an overlap-free layout that has not moved any node (\ie scaling ratio of Scaling~\cite{chen2020survey}). \modelname moves the nodes until there is no overlap (\textit{scan line}~\cite{dwyer2005vpsc}) \textit{and} until the scaling ratio is optimal up to a given precision $s_{step}$ (\eg $0.1$). 

We name \textit{pass} a call to the optimization algorithm \sgdgd (see \autoref{sec:sgd2_optim}) and \textit{iteration} a step within \sgdgd (\ie moving all pairs of nodes once). The maximum number of passes over \sgdgd is defined by the binary search maximum depth $s \le \log \left(\frac{s_{max}-1}{s_{step}}\right)$. Since the pass in \sgdgd costs $\mathcal{O}(N^2)$ and that we test if there remains any overlap right after with the \textit{scan line}~\cite{dwyer2005vpsc} algorithm that executes in $\mathcal{O}(N \log N)$; the final complexity of \modelname is $\mathcal{O}(s(N^2 + N \log N))$.

As every pass in the optimization algorithm resets the annealing step size schedule (see \autoref{sec:sgd2_optim}), the nodes can be moved a lot at the beginning of every pass; meaning that \modelname is somewhat allowed to modify the initial layout more than expected. Hence, we also experiment a variant of \modelname, called \modelnamep, in which the model starts from the scaled initial layout at every pass of the optimization algorithm. We expect this variant to be able to preserve the initial layout even better, probably at the cost of convergence speed.

\autoref{alg:framework} presents \modelname algorithm pseudo-code. It has been simplified by removing the first pass that occurs with the initial scaling of the layout if the sum of nodes areas is lower than the layout bounding box. In practice, it only enters the while loop (line $20$) to search for the optimal scale \textit{if} scaling is necessary. The only modification required to implement \modelnamep is to change line $21$ into $X' \leftarrow \textrm{scaleLayout}(X, curScale)$ so that the next pass in the optimization algorithm starts with the scaled initial layout.

\begin{algorithm}[!h]    
    \begin{algorithmic}[1]
        \caption{\modelname pseudo-code}
        \label{alg:framework}
        \Methods
            \State getScalingRatio($Layout$, $Sizes$): returns the minimum scaling ratio~\cite{chen2019surveyShort,chen2020survey} so that there is no overlap anymore in $Layout$
            \State containsOverlap($Layout$, $Sizes$):~\cite{dwyer2005vpsc} returns $true$ if there is overlap in $Layout$, $false$ otherwise
            \State scaleLayout($Layout$, $scaleFactor$): returns $Layout$ scaled by $scaleFactor$
            \State SGD2\_StressOPT($Layout'$, $Layout$, $Sizes$): One pass of \sgdgd~\cite{zheng2018sgdgd} to optimize overlap removal modeled as stress
        \EndMethods \\
        
        \Variables
            \State $X$: Initial layout, 2D position of every node
            \State $S$: Nodes sizes in 2D, $S_i = (w_i, h_i)$ 
            \State $scaleStep$: Scaling step size to stop the search of optimal scale
        \EndVariables \\
        
        \Procedure{\modelname}{$X$, $S$, $scaleStep$, $SGD2\_HP$}
            \State $lowScale$ $\leftarrow$ $1$
            \State $upScale$ $\leftarrow$ getScalingRatio($X$, $S$)
            \State $curScale$ $\leftarrow$ $(lowScale + upScale)/2$
            \State $X'$ $\leftarrow$ $X$
            \State $thereIsOverlap$ $\leftarrow$ containsOverlap($X'$, $S$) 
            
            \While{$thereIsOverlap$ \OR ($upScale - lowScale > scaleStep$)}
                \State $X'$ $\leftarrow$ scaleLayout($X'$, $curScale$)
                \State $X'$ $\leftarrow$ SGD2\_StressOPT($X'$, $X$, $S$)
                \State $thereIsOverlap$ $\leftarrow$ containsOverlap($X'$, $S$) 
                \If{$thereIsOverlap$}
                    \State $lowScale$ $\leftarrow$ $curScale$
                \Else
                    \State $upScale$ $\leftarrow$ $curScale$
                \EndIf
                    \State $curScale$ $\leftarrow$ $(lowScale + upScale)/2$
            \EndWhile
            \State \Return $X'$
        \EndProcedure
    \end{algorithmic}
\end{algorithm}

\section{\modelname Evaluation}
\label{sec:benchmark}
For the sake of reproducibility, we used the same evaluation protocol as in the Chen \textit{et al.} survey~\cite{chen2019surveyShort,chen2020survey}. That includes quality metrics, datasets and algorithms. This section describes this protocol and presents the results of \modelname comparison with the selected algorithms on these datasets and metrics. Finally, \modelname execution time is also compared with some algorithms.  

\subsection{Evaluation Protocol}

\subsubsection{Quality Metrics.}
\label{sec:quality_metrics}

The quality metrics used to compare \modelname with other algorithms from the literature were selected by Chen~\etal{chen2019surveyShort,chen2020survey}. All are oriented as \textit{lower is better}, the optimal value being 0 unless specified otherwise.

\label{met:oonni}
\textbf{\textit{oo\_nni}}: stands for the \textit{Orthogonal Ordering: Normalized Number of Inversions} and counts the number of times the nodes orthogonal order have been violated.

\label{met:spcha}
\textbf{\textit{sp\_ch\_a}}~\cite{strobelt2012rwordle}: is for \textit{Spread Minimization: Convex Hull Area}. It measures by how much the convex hull area of the overlap-free layout is different from the one of the initial layout: $sp\_ch\_a = \frac{\textrm{convex\_hull\_area}(X')}{\textrm{convex\_hull\_area}(X)}$, the optimal value being $1$. This metric mainly measures the layout scaling.

\label{met:gsbbiar}
\textbf{\textit{gs\_bb\_iar}}: means \textit{Global Shape preservation: Bounding Box Improved Aspect Ratio} and is a variant of the aspect ratio between the bounding box of the initial and overlap-free layouts in which the minimal and target value is 1.

\label{met:dmimse}
\textbf{\textit{nm\_dm\_imse}}: stands for \textit{Node Movement minimization: Distance Moved Improved Mean Squared Error}. It quantifies how much the nodes moved from their position in the initial layout to theirs in the overlap-free layout. To lessen the effect of positions value domains, the layouts are aligned as follows:
\begin{equation}
    \textrm{nm\_dm\_imse} = \frac{1}{N} \sum_{v_i \in V} ||X'_i - \textrm{scale}(\textrm{shift}(X_i))||^2
\end{equation}
where \textit{shift} and \textit{scale} are moving and scaling the initial layout bounding box to match the center and dimensions of the overlap-free layout.
It is important to mention that both in \cite{chen2019surveyShort,chen2020survey} and therefore in this paper, the bounding boxes computed to scale the layouts do not take the nodes sizes into account.

\label{met:elrsd}
\textbf{\textit{el\_rsdd}}: is for \textit{Edge Length preservation: Relative Standard Deviation Delaunay}. It measures by how much the lengths of edges in the Delaunay Triangulation graph of an initial layout are preserved, \ie how well short-distances are preserved. 

\subsubsection{Datasets.}
\label{sec:datasets}

Still following Chen~\etal{chen2019surveyShort,chen2020survey} evaluation protocol, we use the Generated and Graphviz datasets available online\footnote{Generated and Graphviz graphs: \url{https://github.com/agorajs/agora-dataset},\\last consulted on May 2022}.

\textbf{Generated} is a set of \num{840} synthetic graphs specifically generated for the benchmark in \cite{chen2019surveyShort,chen2020survey}. It is made of $120$ graphs of each size ${10, 20, 50, 100, 200, 500, 1000}$, laid out with the $FM^3$ algorithm~\cite{hachul2004fm3}. These layouts have $2770 \pm 7567 (std)$ initial overlaps in average, ranging between $0$ and $31843$.

\textbf{Graphviz} is a set of \num{14} real-world graphs from the Graphviz suite. They have between $36$ and $1463$ nodes and are laid out with $SFDP$ algorithm~\cite{hu2005sfdp} and have for between $4$ and $11582$ initial overlaps ($2118\pm4078(std)$ in average).

\subsubsection{Baseline Algorithms.}
\label{sec:sota_algo}

As already stated (see \autoref{sec:related_works}), there are two main criteria to optimize in overlap removal algorithms: compactness and initial layout preservation. By design, \modelname belongs to the second category and is then only compared with its corresponding algorithms (\ie PFS~\cite{misue1995pfs}, PFS'~\cite{hayashi1998pfsprime}, PRISM~\cite{gansner2010prism}, GTREE~\cite{nachmanson2016gtree} and Diamond~\cite{meulemans2019diamond}). Other algorithms create embeddings so compact that it is not even possible to visualize the graph edges and structures anymore; meaning they are not suited to overlap removal for graph visualization. \textit{Scaling} is also excluded since it does not look for balance and rawly upscales the layout; which is not a satisfactory solution on its own.

\subsection{Comparison with Baseline Algorithms on Quality Metrics}
\label{sec:quantitative_bench}

This section reports and discusses the performances of \modelname and the selected algorithms from the literature on the \textbf{Generated} and \textbf{Graphviz} datasets.

\begin{figure}[!b]
    \centering
    \includegraphics[width=\linewidth]{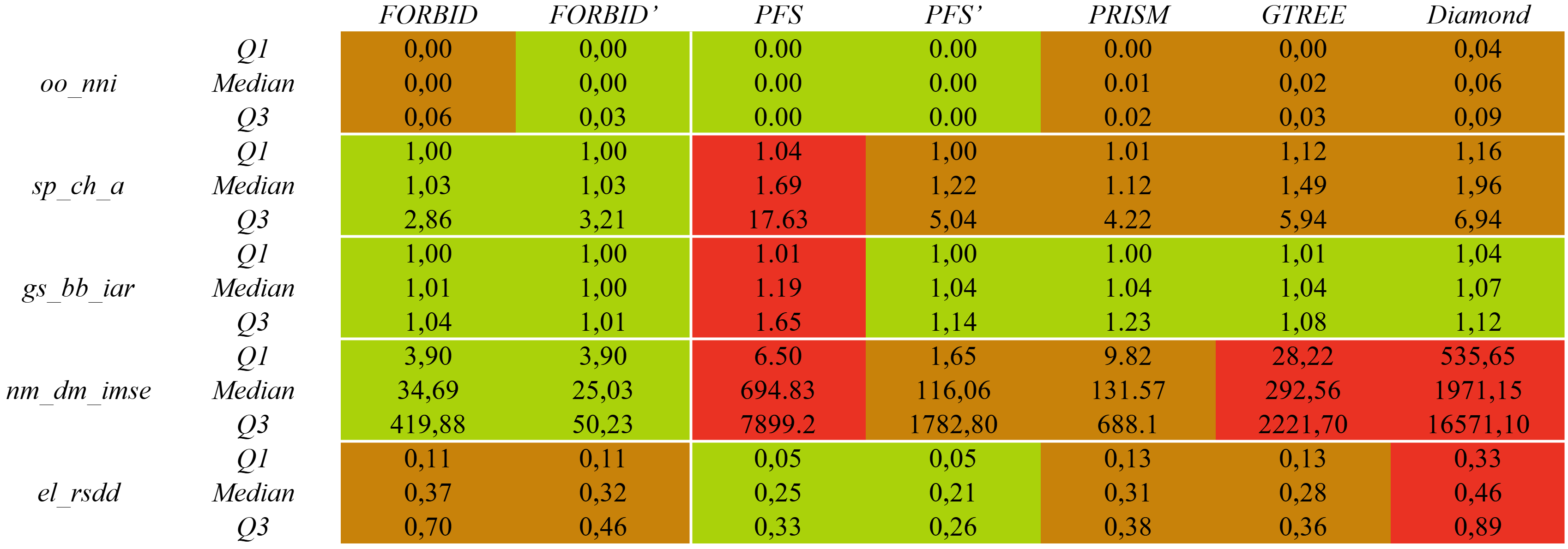}
    \caption{Quality metrics quartile values for each algorithm on the \textbf{Generated} dataset (see \custRef{sec:datasets}{Sect.}). Cells color are selected based on the median value of the algorithms on each metric to enhance comparisons readability. The greener/lighter a cell color is, the better its quality metric score.}
    \label{fig:quality_generated}
\end{figure}

\begin{figure}[!b]
    \centering
    \includegraphics[width=\linewidth]{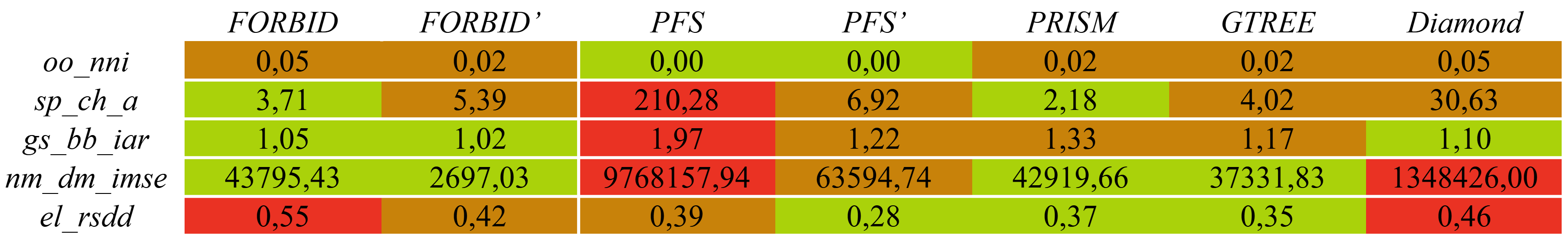}
    \caption{Quality metrics mean values on the \textbf{Graphviz} dataset (see \custRef{sec:datasets}{Sect.}). The greener/lighter a cell color is, the better its quality metric score.}
    \label{fig:quality_real}
\end{figure}

On the \textbf{Generated} dataset (see \autoref{fig:quality_generated}), every algorithm succeeds in minimizing \oonni, with Diamond and \modelname having slightly higher scores than others on worst cases (Q3). On \spcha and \gsbbiar, \modelname and \modelnamep have the best scores by a fair margin, especially on worst cases (Q3). This demonstrates a good capability to limit the upscaling of the drawing in complicated layouts. Both \modelname and \modelnamep also minimize nodes movements \dmimse more than other algorithms, especially \modelnamep that barely moves the nodes even on complex cases (\dmimse$=50.23$ on Q3). 
Finally, \modelname and its variant both have higher scores on \elrsd. As defined in \autoref{sec:quality_metrics}, this metric measures the edge length preservation along the edges of the Delaunay Triangulation (DT) of a graph. Hence, it does not measures that the lengths of the actual graph edges are preserved, but rather quantifies the preservation of the distances between the closest nodes. On the other hand, \modelname focuses on the preservation of \textit{all} the node pairwise distances. These two strategies do not have the same notion of \textit{preservation} of the initial layout and by ignoring long distances, PRISM tends to break the overall layout aspect. In addition, since overlapped nodes are likely to be adjacent in the DT graph and since our ideal distance between them is not the shortest possible (see \autoref{sec:stress_model}), it was expected that \modelname would distord these distances more than other algorithms (\eg PRISM).

The performances on the \textbf{Graphviz} dataset are reported in \autoref{fig:quality_real}. On that dataset, the same trend can be observed with slight differences. Here again, every algorithm successfully minimizes \oonni and both \modelname and \modelnamep are still ahead by a fair margin on \gsbbiar. 
However, they are no longer the bests on \spcha and \modelnamep is the only one to keep the lead on \dmimse. On those two metrics, PRISM is slightly better than \modelname, while \modelnamep has a deteriorated \spcha for a much better node movement \dmimse in comparison to other algorithms and to its own performances on the \textbf{Generated} dataset. In fact, since \modelnamep focuses on preserving the initial layout, it tends to upscale the layouts more (worsening \spcha) to minimize the nodes movements (improving \dmimse). Finally, both have relatively high \elrsd scores.

\subsection{Execution Time Comparison with some Baseline Algorithms}
\label{sec:exec_time}

This section compares \modelname execution time with that of the best techniques according to the performances observed in \autoref{sec:quantitative_bench}: PFS', GTREE and PRISM.

\begin{figure}[!b]
    \centering
    \includegraphics[width=\linewidth]{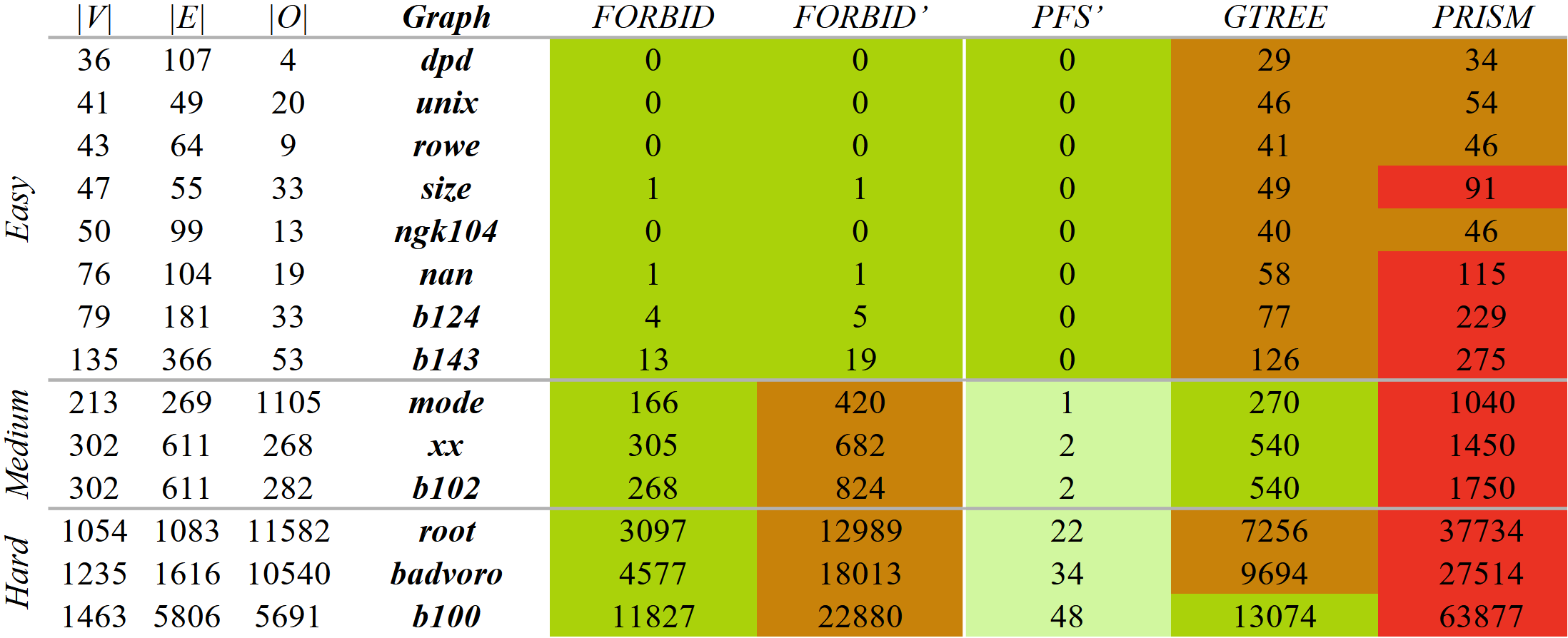}
    \caption[toto]{Execution time (in ms) of \modelname, \modelnamep, GTREE and PRISM on the \textbf{Graphviz} dataset (see \custRef{sec:datasets}{Sect.}). Rows are sorted by ``graph complexity'' measured with their number of nodes \CardV, edges \CardE and overlaps $|O|$. The greener/lighter a cell color is, the better its quality metric score.}
    \label{fig:exectime_real}
\end{figure}

Execution times on the \textbf{Graphviz} dataset are reported in \autoref{fig:exectime_real} and enable to categorize three groups of difficulty. \modelname, \modelnamep and PFS' are instantaneous on \textit{easy} graphs, taking less than $20ms$ to solve the OR problem, being about ten to twenty times faster than GTREE and PRISM. 
On \textit{medium} complexity graphs, PFS' achieves again the best performances. \modelname is slower, but remains faster than GTREE and PRISM by a fair margin, while \modelnamep is slower than GTREE. On these graphs, PRISM is significantly slower than the four others. 
Finally, on the \textit{hard}est graphs, PFS' is still almost instantaneous. \modelname is faster than the remaining methods, while \modelnamep loses to GTREE. On \textit{hard} graphs, PRISM becomes dramatically slower than other algorithms. 

In the end, we can conclude that \modelname scales well with the OR problem complexity, while it is more difficult for \modelnamep. For this last, the constraint to start from the scaled initial layout at each pass in the optimization algorithm significantly slows its convergence (see \autoref{sec:convergence}) and is responsible of the higher upscaling observed in \autoref{sec:quantitative_bench} that enables a better layout preservation. In comparison with PRISM, which also optimizes OR modeled as stress, both \modelname variants are much faster, while PFS' faster handles large graphs.

\section{Discussion}
\label{sec:discussion}
This section discusses \modelname behavior on some specific graph examples by comparing it to other algorithms. It also briefly discusses the convergence of \modelname and \modelnamep variants, highlighting their different behaviors. 

\subsubsection{Visual Evaluation.}
\label{sec:visual_eval}

\begin{figure}
    \centering
    \begin{tabular}{cccc}
        & \textit{mode} & \textit{badvoro} & \textit{root} \\

        \begin{sideways}{\hspace{0.2cm}\textbf{Initial Layout}}\end{sideways}&
        \includegraphics[width=.300\linewidth]{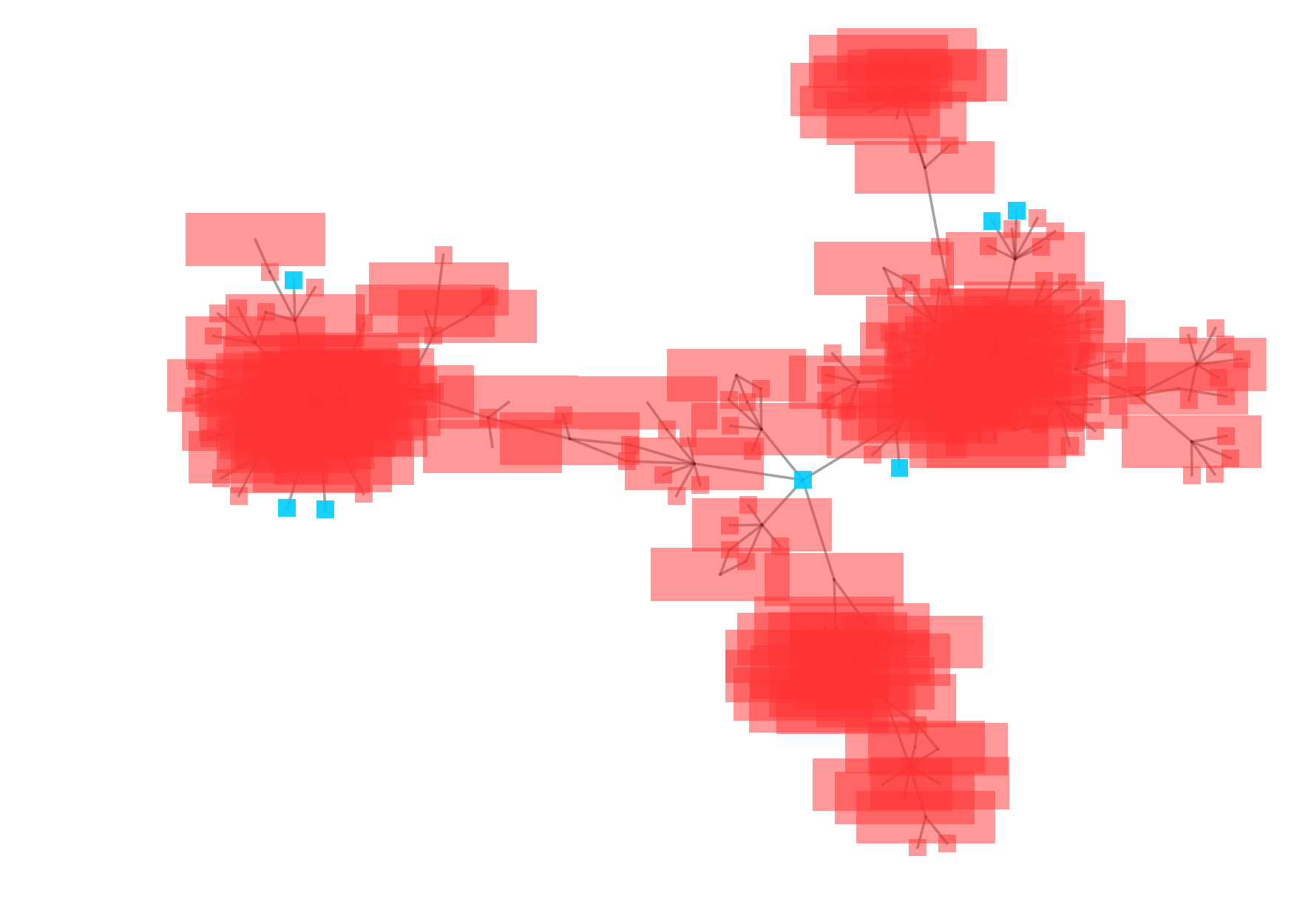} & 
        \includegraphics[width=.300\linewidth]{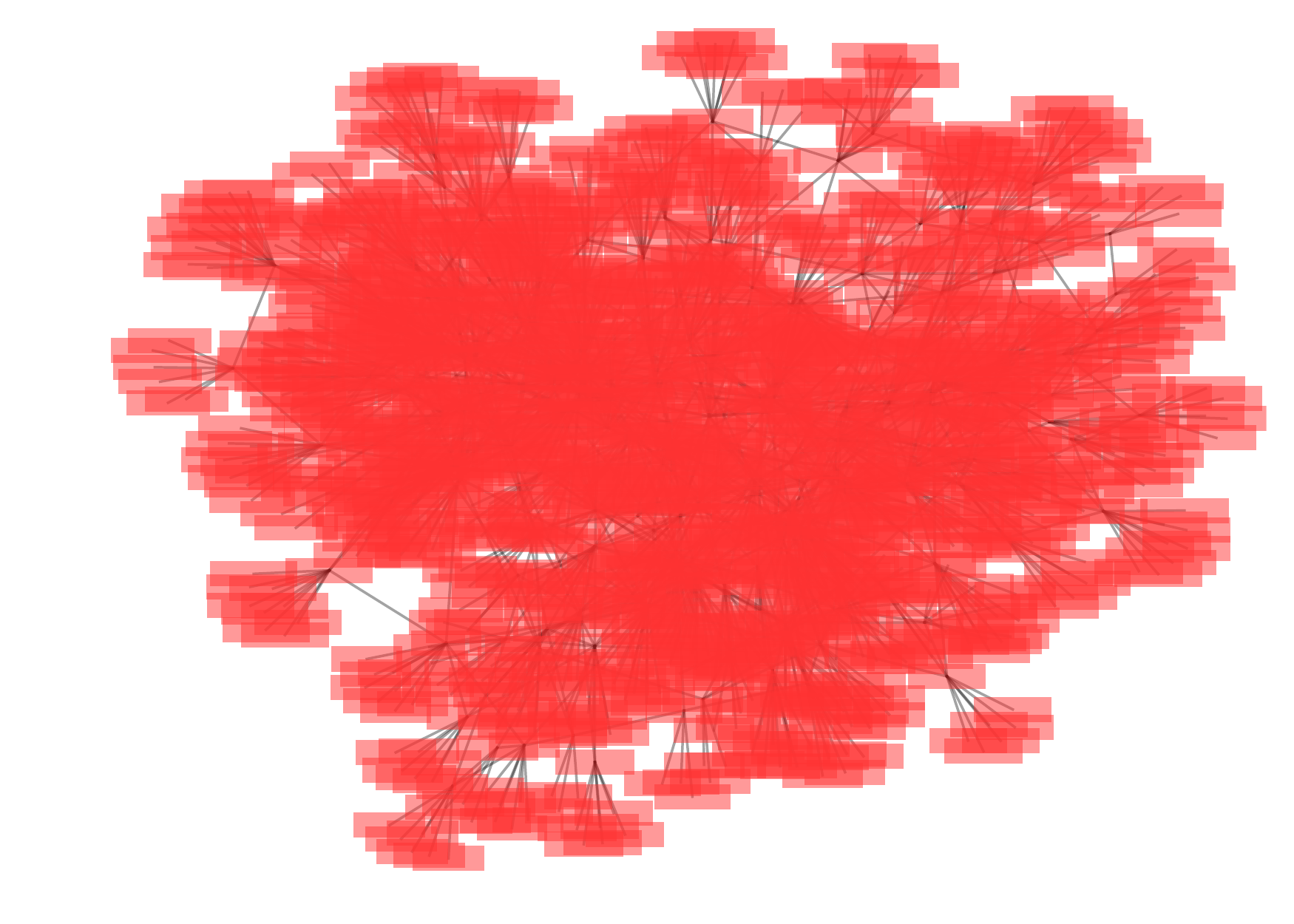} &
        \includegraphics[width=.300\linewidth]{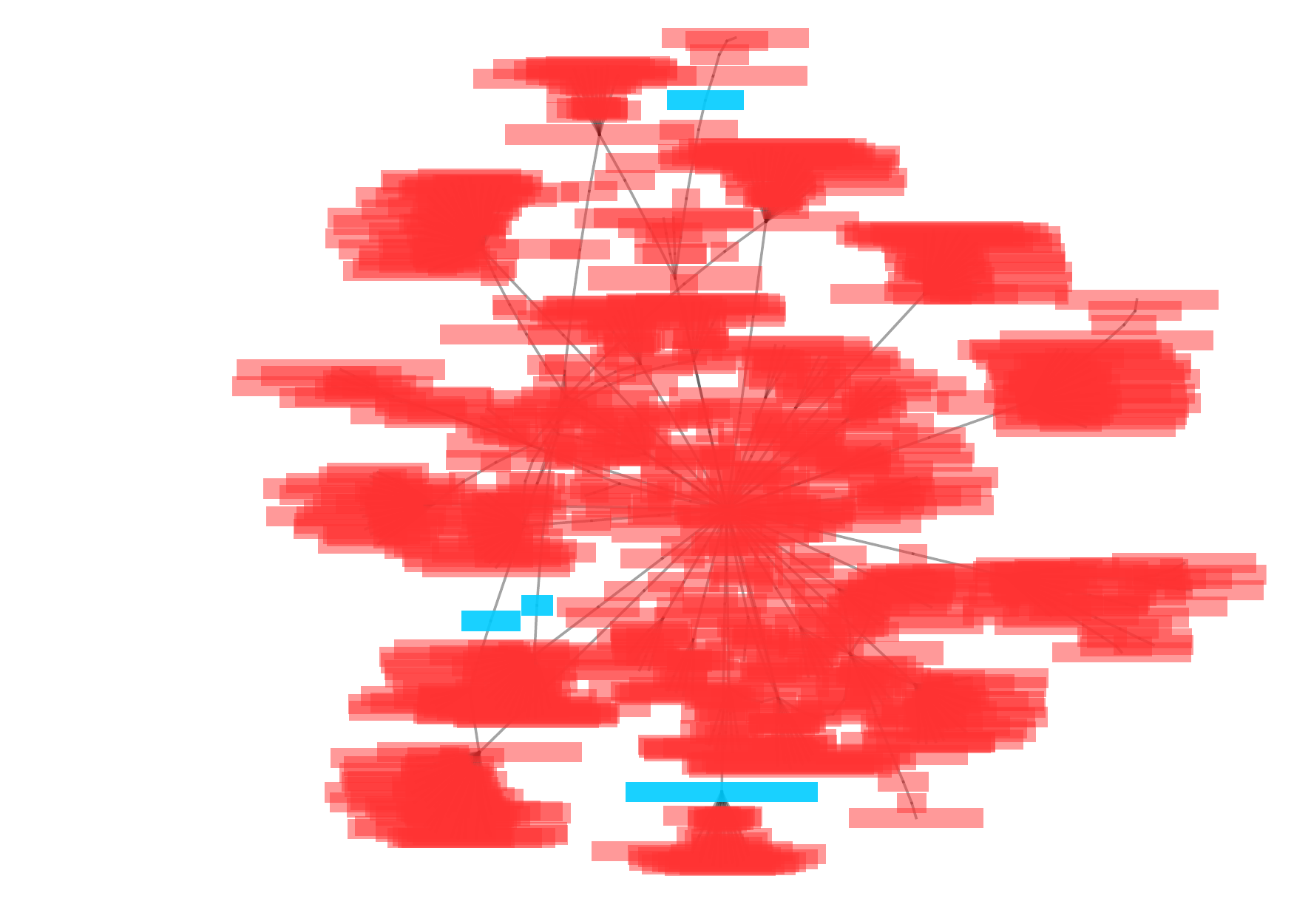} \\

        \begin{sideways}{\hspace{0.5cm}\textbf{\modelname}}\end{sideways}&
        \includegraphics[width=.300\linewidth]{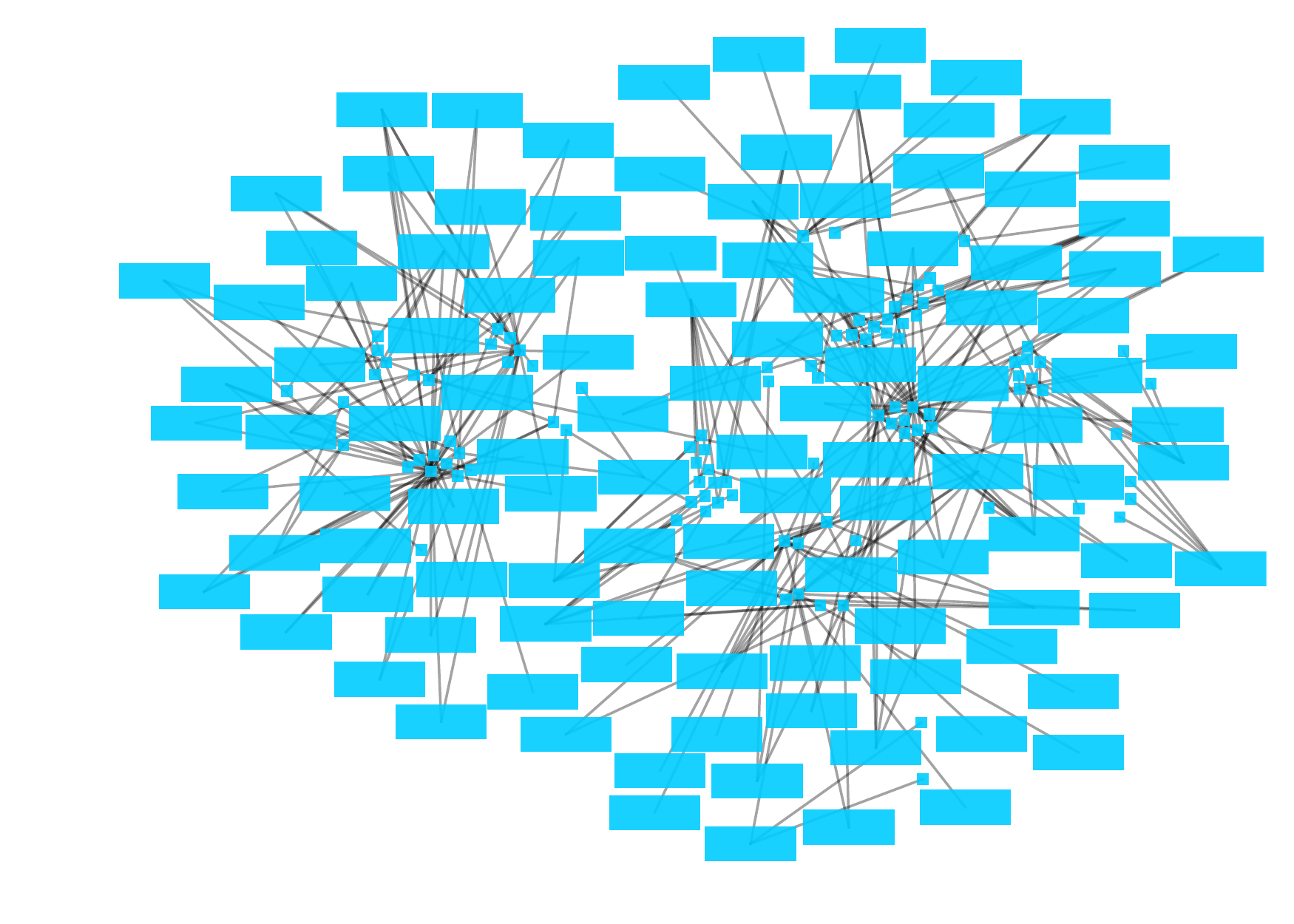} & 
        \includegraphics[width=.300\linewidth]{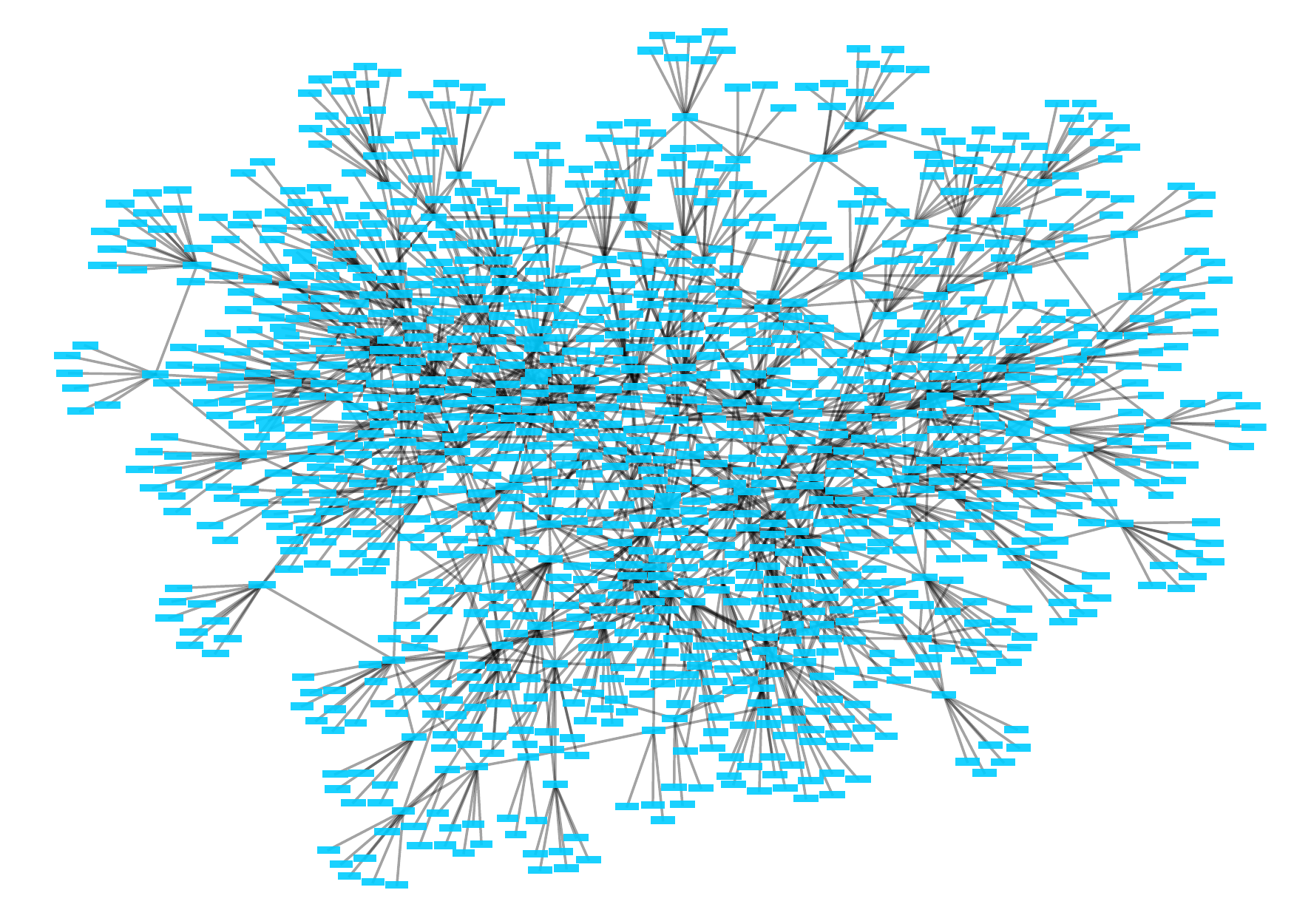} &
        \includegraphics[width=.300\linewidth]{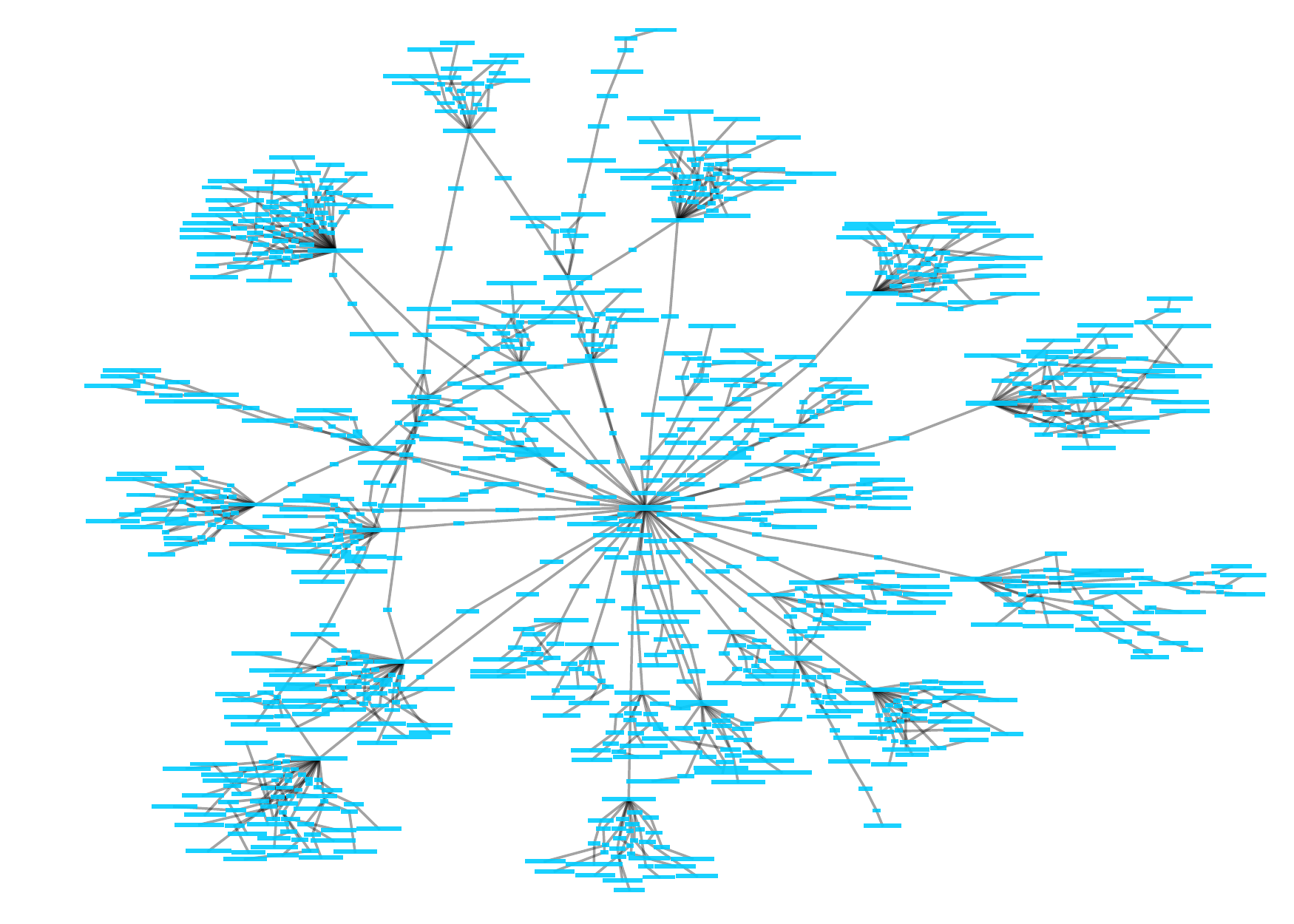} \\

        \begin{sideways}{\hspace{0.5cm}\textbf{\modelnamep}}\end{sideways}&
        \includegraphics[width=.300\linewidth]{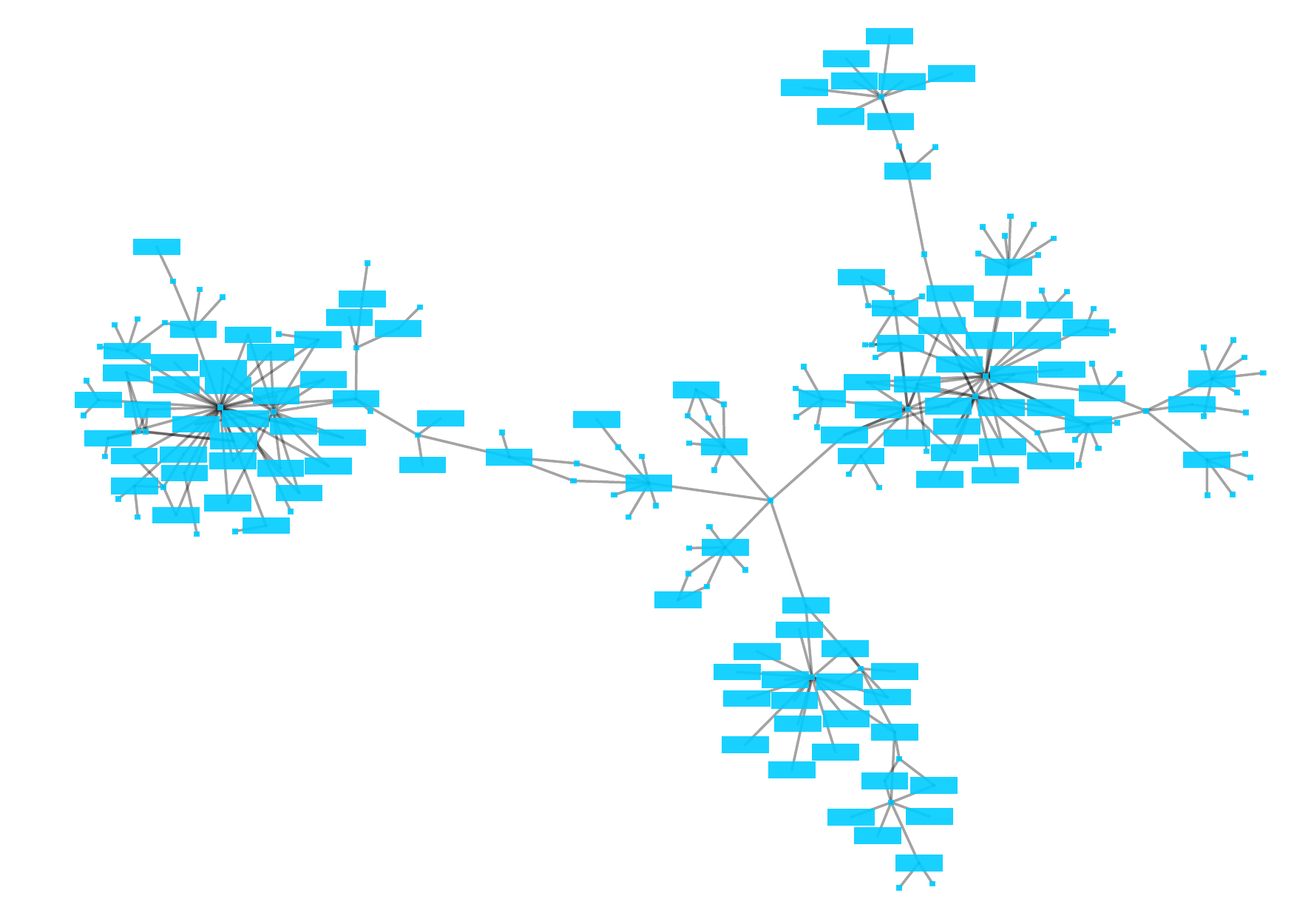} & 
        \includegraphics[width=.300\linewidth]{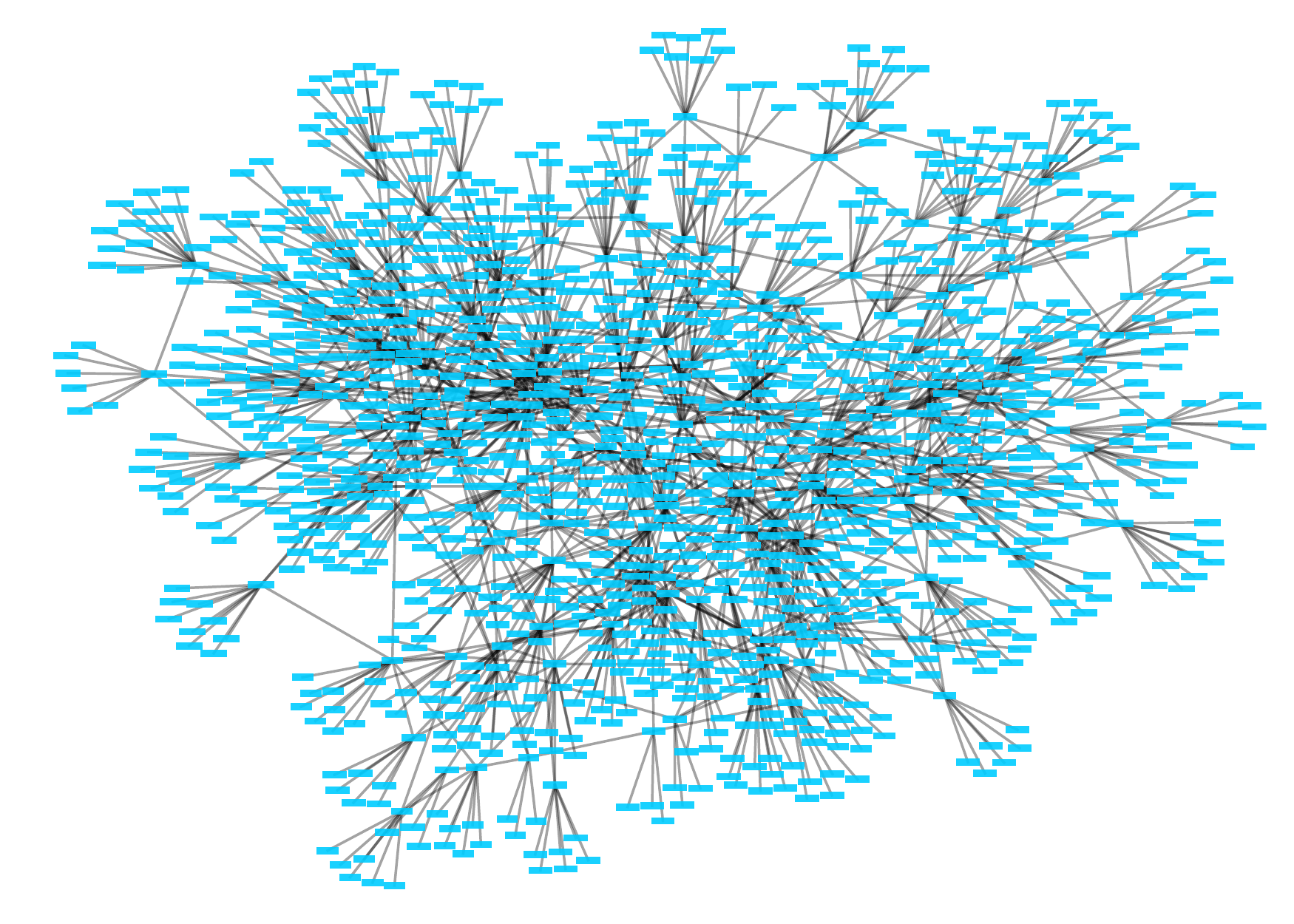} &
        \includegraphics[width=.300\linewidth]{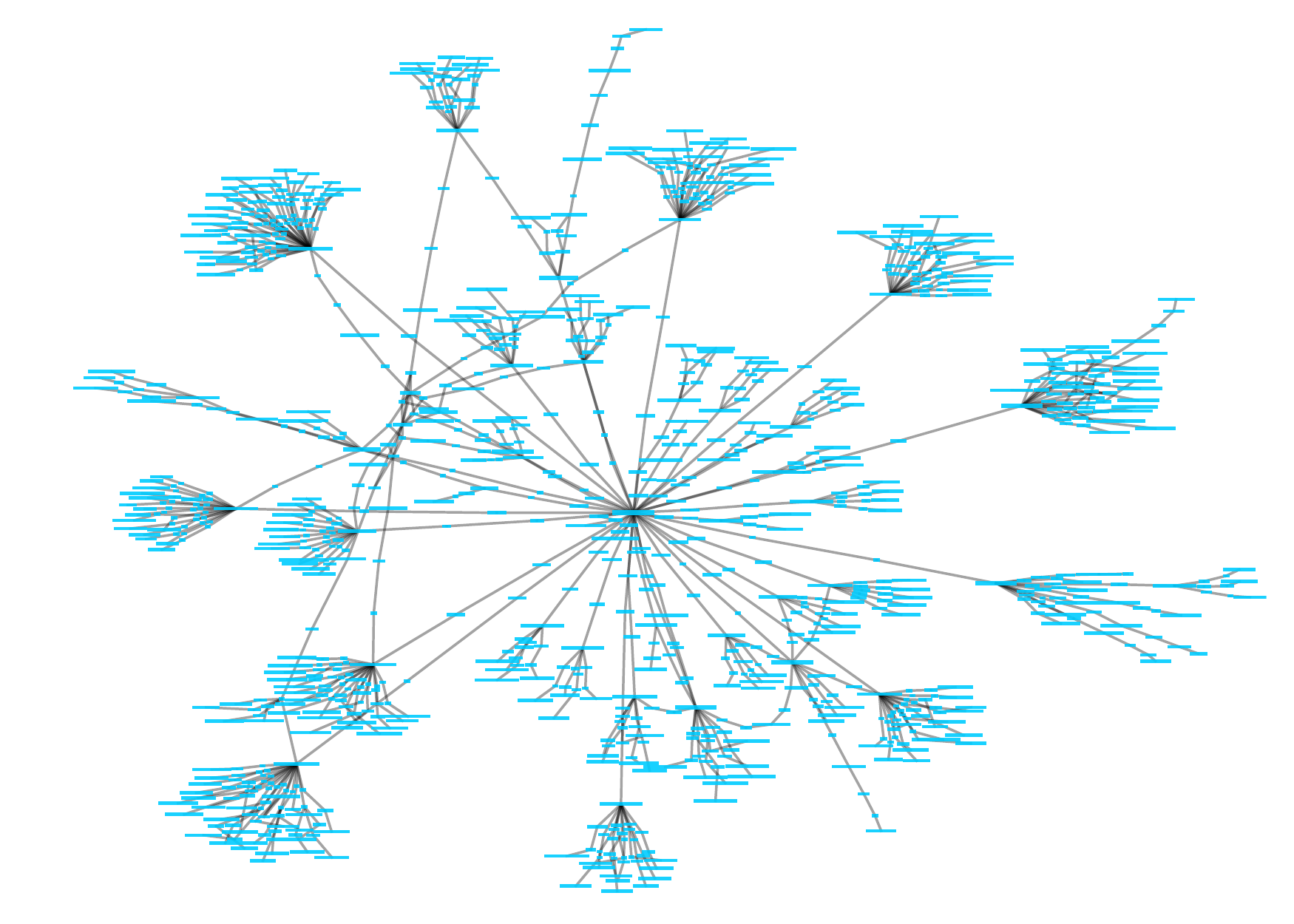} \\

        \begin{sideways}{\hspace{1cm}\textbf{PFS'}}\end{sideways}&
        \includegraphics[width=.300\linewidth]{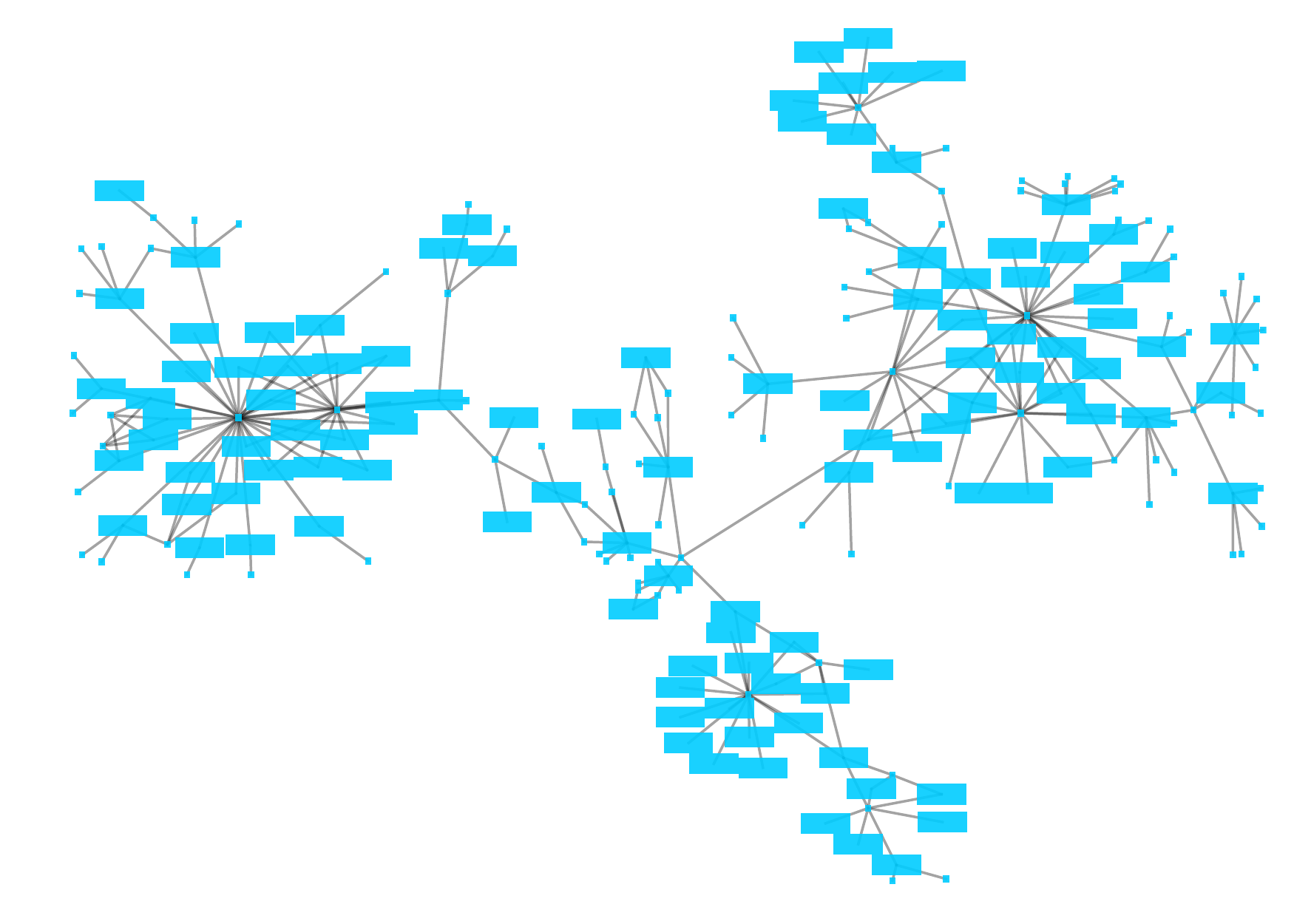} & 
        \includegraphics[width=.300\linewidth]{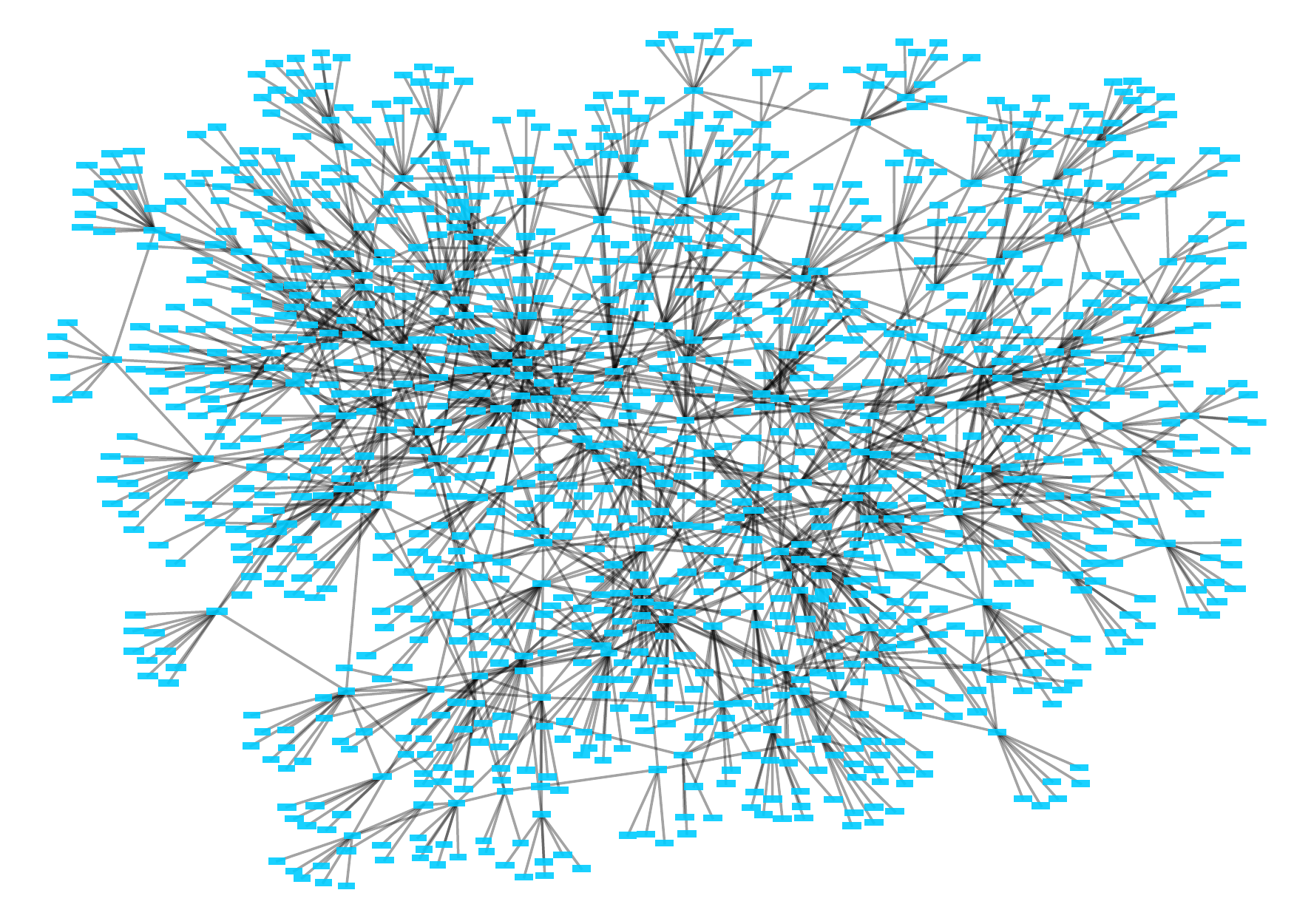} &
        \includegraphics[width=.300\linewidth]{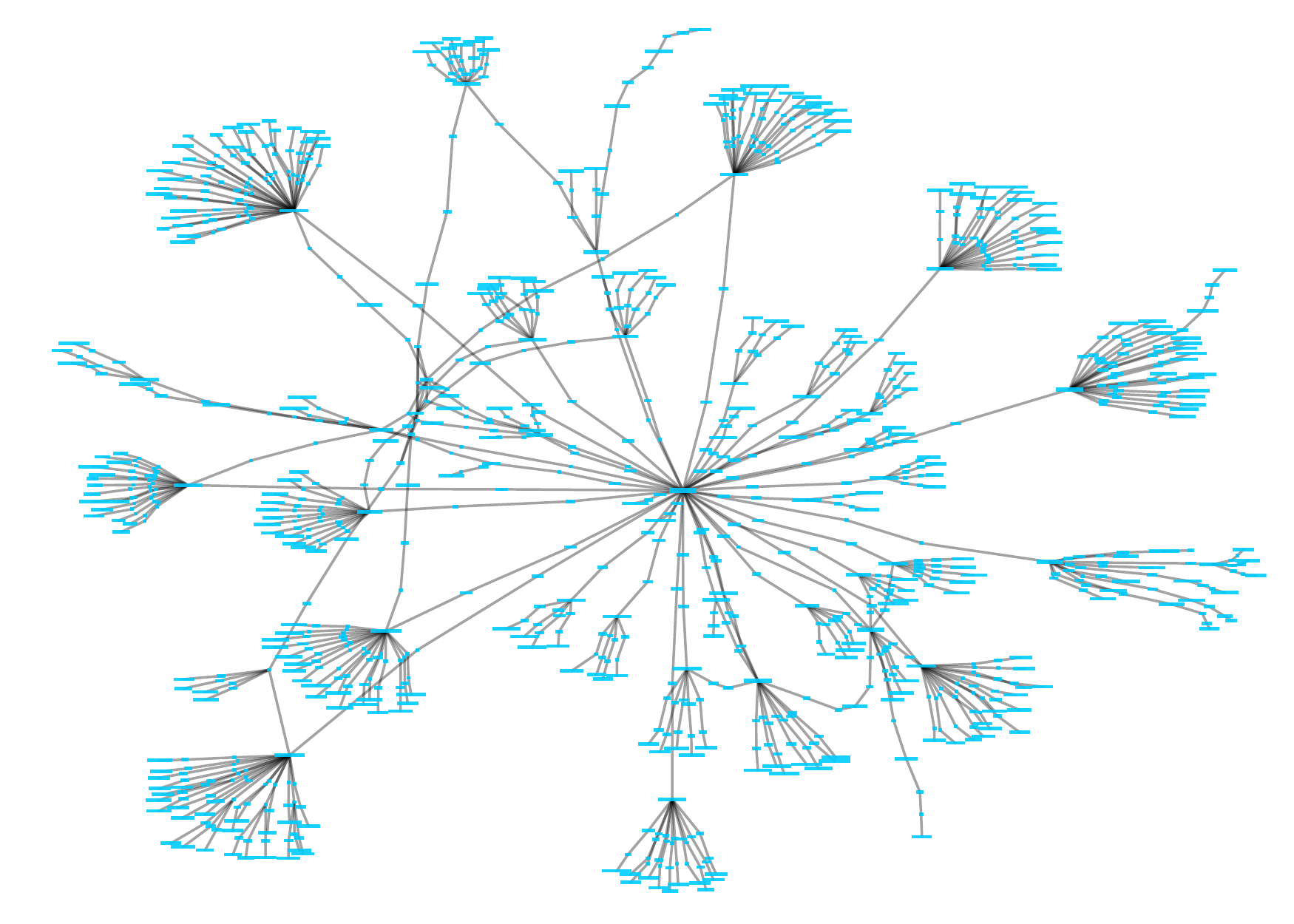} \\

        \begin{sideways}{\hspace{0.5cm}\textbf{GTREE}}\end{sideways}&
        \includegraphics[width=.300\linewidth]{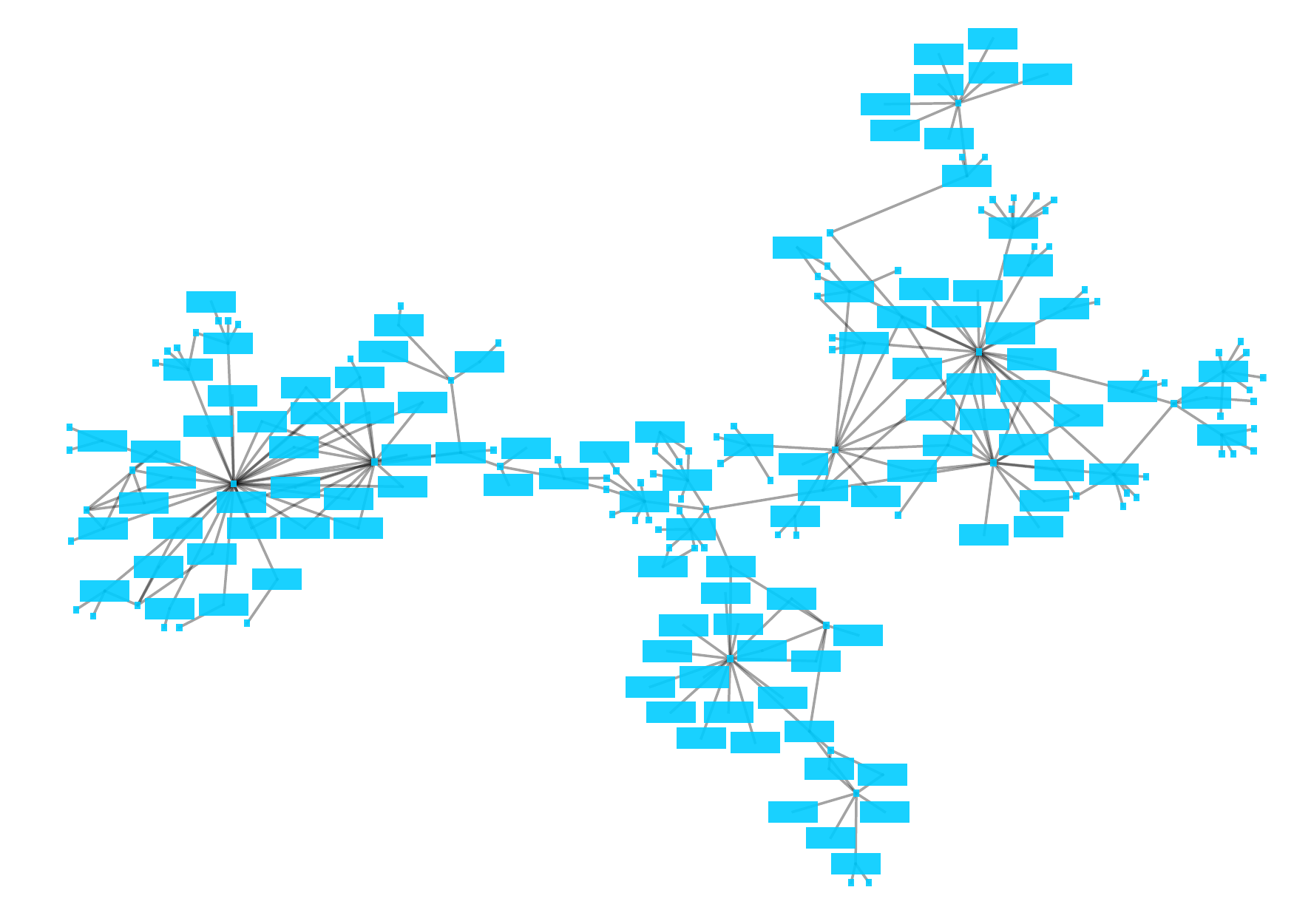} & 
        \includegraphics[width=.300\linewidth]{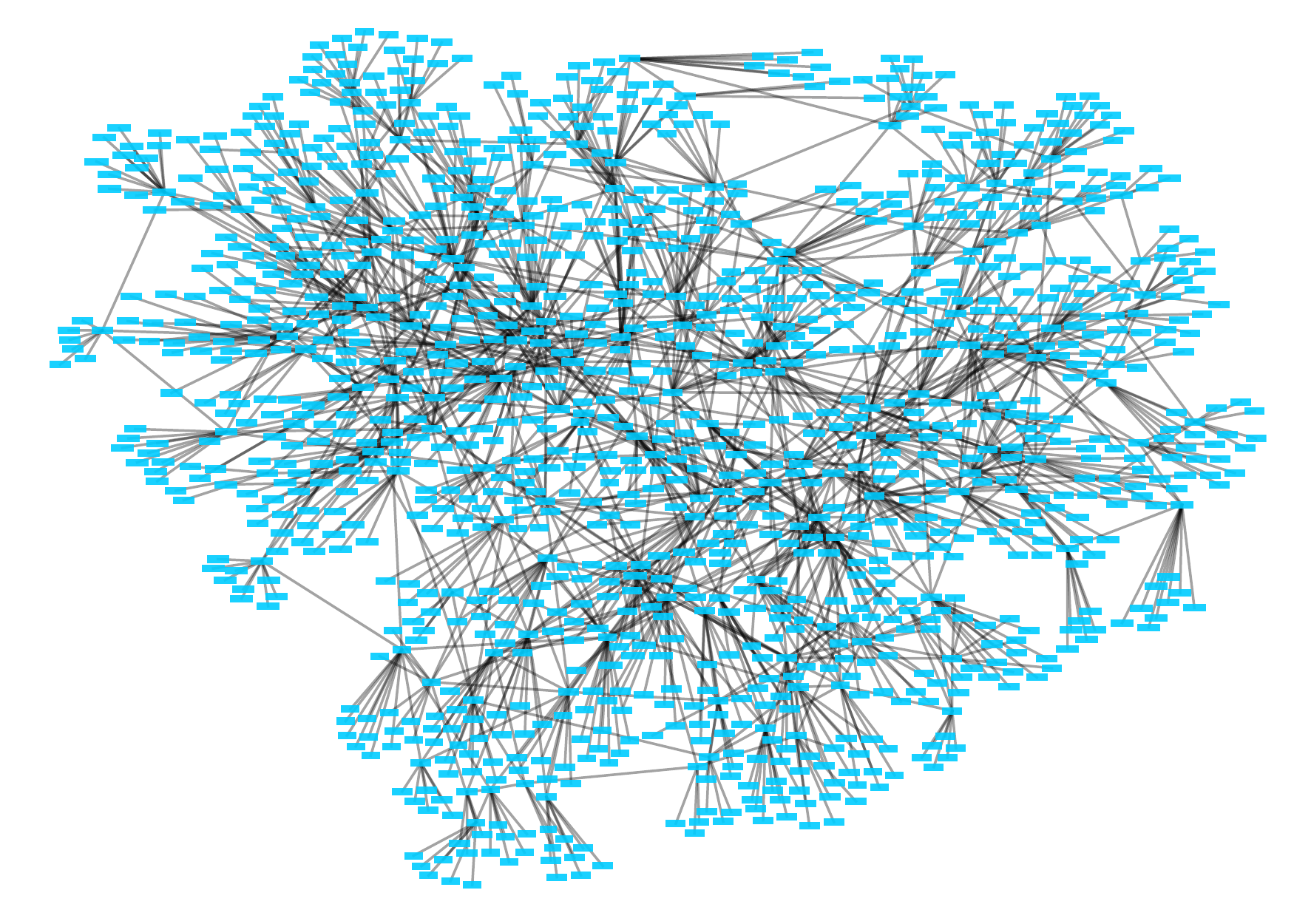} &
        \includegraphics[width=.300\linewidth]{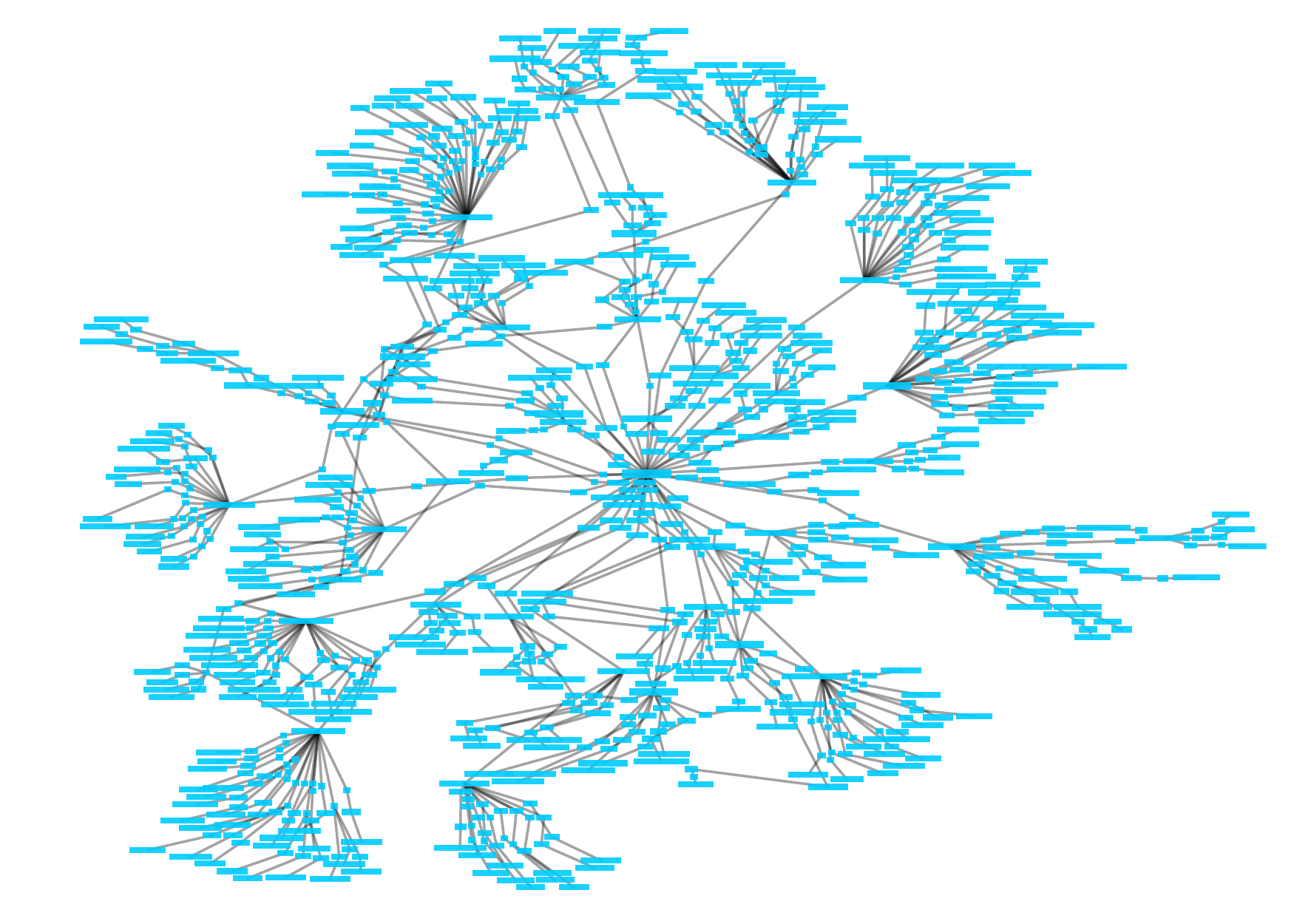} \\

        \begin{sideways}{\hspace{0.5cm}\textbf{PRISM}}\end{sideways}&
        \includegraphics[width=.300\linewidth]{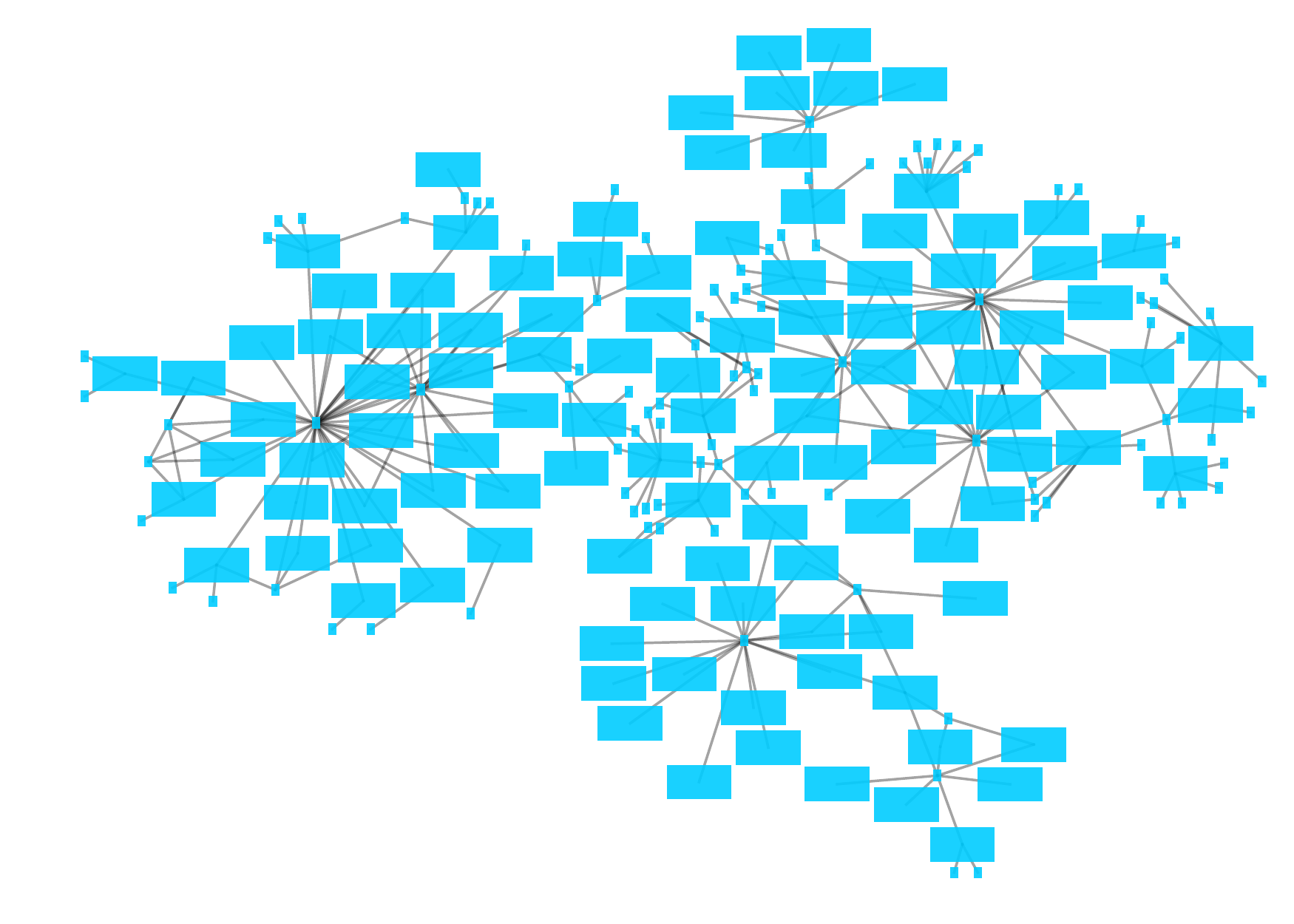} & 
        \includegraphics[width=.300\linewidth]{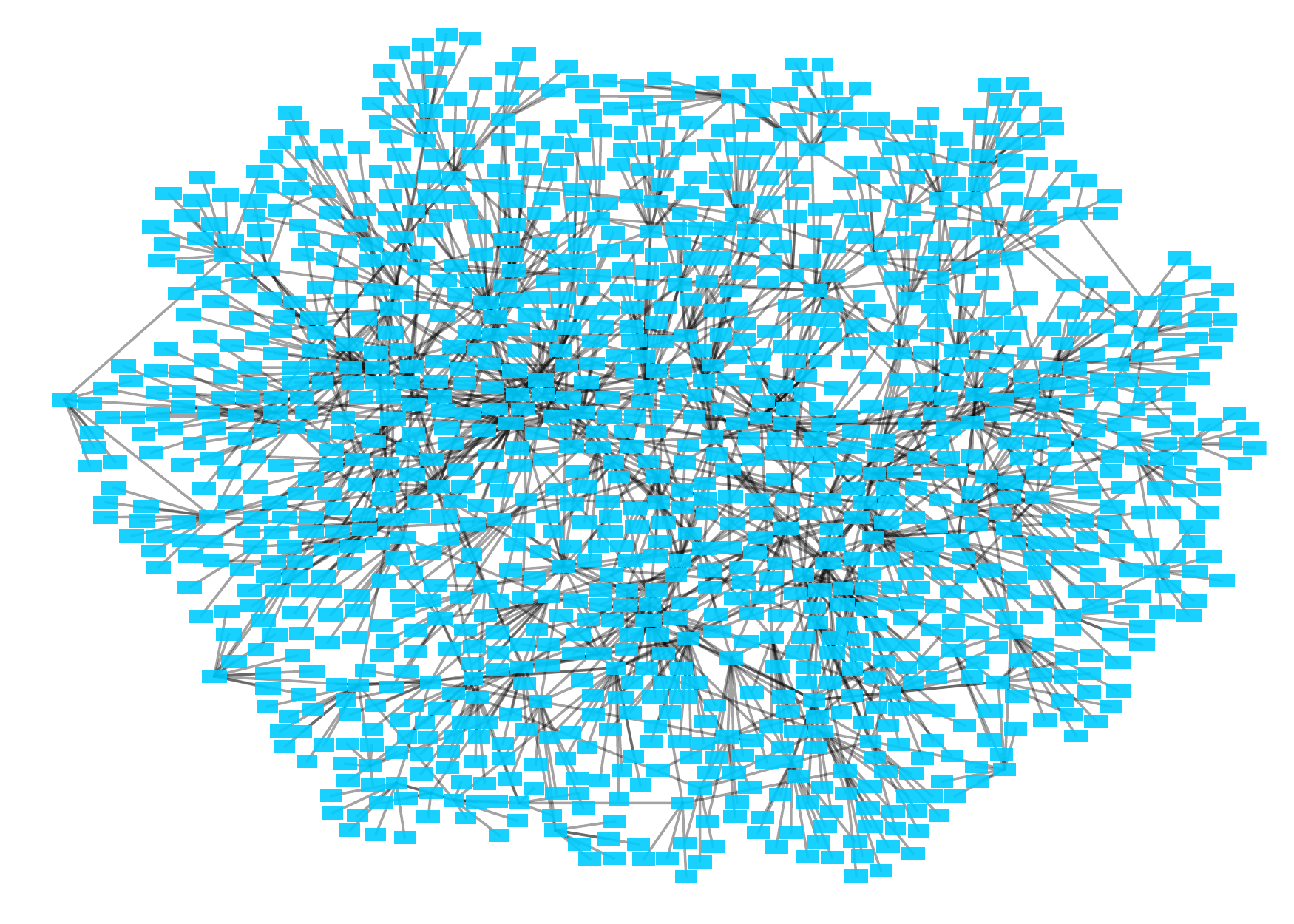} &
        \includegraphics[width=.300\linewidth]{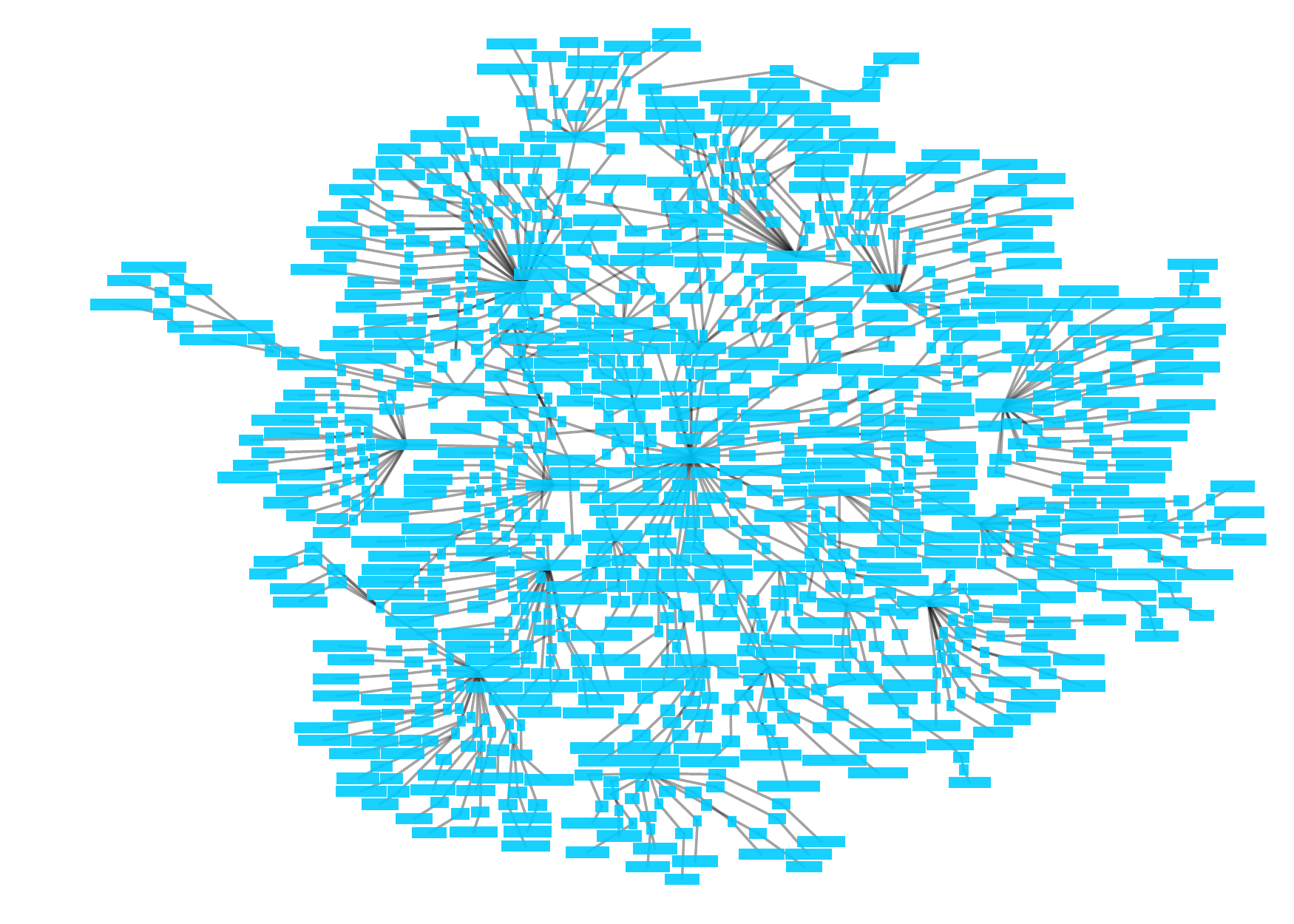} \\
        
    \end{tabular}
    \caption{Graph visualization of examples from the \textbf{Graphviz} dataset with \modelname, \modelnamep, PFS' GTREE and PRISM. Nodes are colored in transparent red if they overlap each other, and opaque blue otherwise.}
    \label{fig:visual_eval}
\end{figure}

This section focuses on the study of \modelname, \modelnamep, PFS', GTREE and PRISM on three graphs of the \textbf{Graphviz} dataset: \textit{mode}, \textit{badvoro} and \textit{root}. Both initial and overlap-free layouts are presented in \autoref{fig:visual_eval}, while the methods quality metrics are reported in \autoref{tab:visual_eval_metrics}. 

On \textit{mode}, both \modelname and PRISM damaged the initial layout to produce more compact embeddings. On the other hand, the others preserved the initial layout structure, \modelnamep being the most pleasing one. The quality metrics corroborate this observation, with \modelname and PRISM having lower \spcha (\ie upscaling) scores while \modelnamep, PFS' and GTREE have smaller \dmimse (\ie nodes movements) scores; \modelnamep having the smallest one.

On both \textit{badvoro} and \textit{root}, the same trend can be observed between the Overlap Removal methods. PRISM produces more compact embeddings that makes it difficult to recognize the initial layout and even to visualize edges. On the other hand, GTREE produces less compact layouts, but distorts the initial graph structures. Finally, \modelname, \modelnamep and PFS' are satisfactory, but PFS' upscales the initial layout more than necessary. Again, these observations are corroborated by the quality metrics presented in \autoref{tab:visual_eval_metrics}: PRISM is consistently better on \spcha, GTREE is not the best on any metric, and the remaining three have less nodes movements but higher upscaling. The high \dmimse scores of PFS' can be imputed to the fact that this metric is sensitive to the overlap-free layout scale, even though the graph layout structures seem visually preserved. 

Overall, \modelname and \modelnamep produce balanced layout that optimize both the initial layout preservation and the embedding compactness.

\begin{table}
    \caption{Quality metrics of the drawings presented in \autoref{fig:visual_eval}. For each graph and quality metric, the best score is highlighted in bold.}
    \vspace{0.1cm}
    \begin{tabular}{|c|cccccc|}
        \hline
        & & \hspace{0.25cm}\textbf{\modelname}\hspace{0.25cm} & \hspace{0.25cm}\textbf{\modelnamep}\hspace{0.25cm} & \hspace{0.25cm}\textbf{PFS'}\hspace{0.25cm} & \hspace{0.25cm}\textbf{GTREE}\hspace{0.25cm} & \hspace{0.25cm}\textbf{PRISM}\hspace{0.25cm}\\
        \hline

        \multirow{5}{*}{\begin{sideways}\textit{mode}\end{sideways}}&
        \textit{oo\_nni}        & 0.20  & \textbf{0.03}  &0.54   & 0.05  &0.06 \\
        &\textit{sp\_ch\_a}     & \textbf{2.98}  & 10.35 &9.08   & 7.77  &3.72 \\
        &\textit{gs\_bb\_iar}   & 1.02  & \textbf{1.01}  &1.20   & 1.23  &1.52 \\
        &\textit{nm\_dm\_imse}  & 20417 & \textbf{1657}  &8655   & 10238 &17442 \\
        &\textit{el\_rsdd}       & 0.95  & 0.70  &\textbf{0.39}   & 0.54  &0.62 \\
        \hline

        \multirow{5}{*}{\begin{sideways}\textit{badvoro}\end{sideways}}&
        \textit{oo\_nni}        & 0.01  & \textbf{0.00}  &0.48   & 0.02  &0.02 \\
        &\textit{sp\_ch\_a}     & 11.42 & 12.40 &17.92  & 13.45 &\textbf{6.87} \\
        &\textit{gs\_bb\_iar}   & \textbf{1.00}  & \textbf{1.00}  &1.25   & 1.17  &1.85 \\
        &\textit{nm\_dm\_imse}  & 979   & \textbf{451}   &57267  & 67296 &47708 \\
        &\textit{el\_rsdd}       & 0.32  & 0.23  &\textbf{0.20}   & 0.40  &0.39 \\
        \hline

        \multirow{5}{*}{\begin{sideways}\textit{root}\end{sideways}}&
        \textit{oo\_nni}        & \textbf{0.01}  & \textbf{0.01}  &0.53   & 0.05  &0.04 \\
        &\textit{sp\_ch\_a}     & 19.17 & 29.99 &48.34  & 14.29 &\textbf{6.2}7 \\
        &\textit{gs\_bb\_iar}   & \textbf{1.01}  & \textbf{1.01}  &1.34   & 1.30  &2.04 \\
        &\textit{nm\_dm\_imse}  & 24502 & \textbf{12151} &662185 & 356632&450287 \\
        &\textit{el\_rsdd}       & 0.90  & 0.72  &\textbf{0.43}   & 0.67  &0.73 \\
        \hline
    \end{tabular}
    \label{tab:visual_eval_metrics}
\end{table}

\subsubsection{Convergence Analysis.}
\label{sec:convergence}

\begin{figure}[!tb]
    \centering
    \begin{tabular}{cccc}
        & \textit{mode} & \textit{badvoro} & \textit{root} \\

        \begin{sideways}{\hspace{1cm}\textbf{\modelname}}\end{sideways}&
        \includegraphics[width=.31\linewidth]{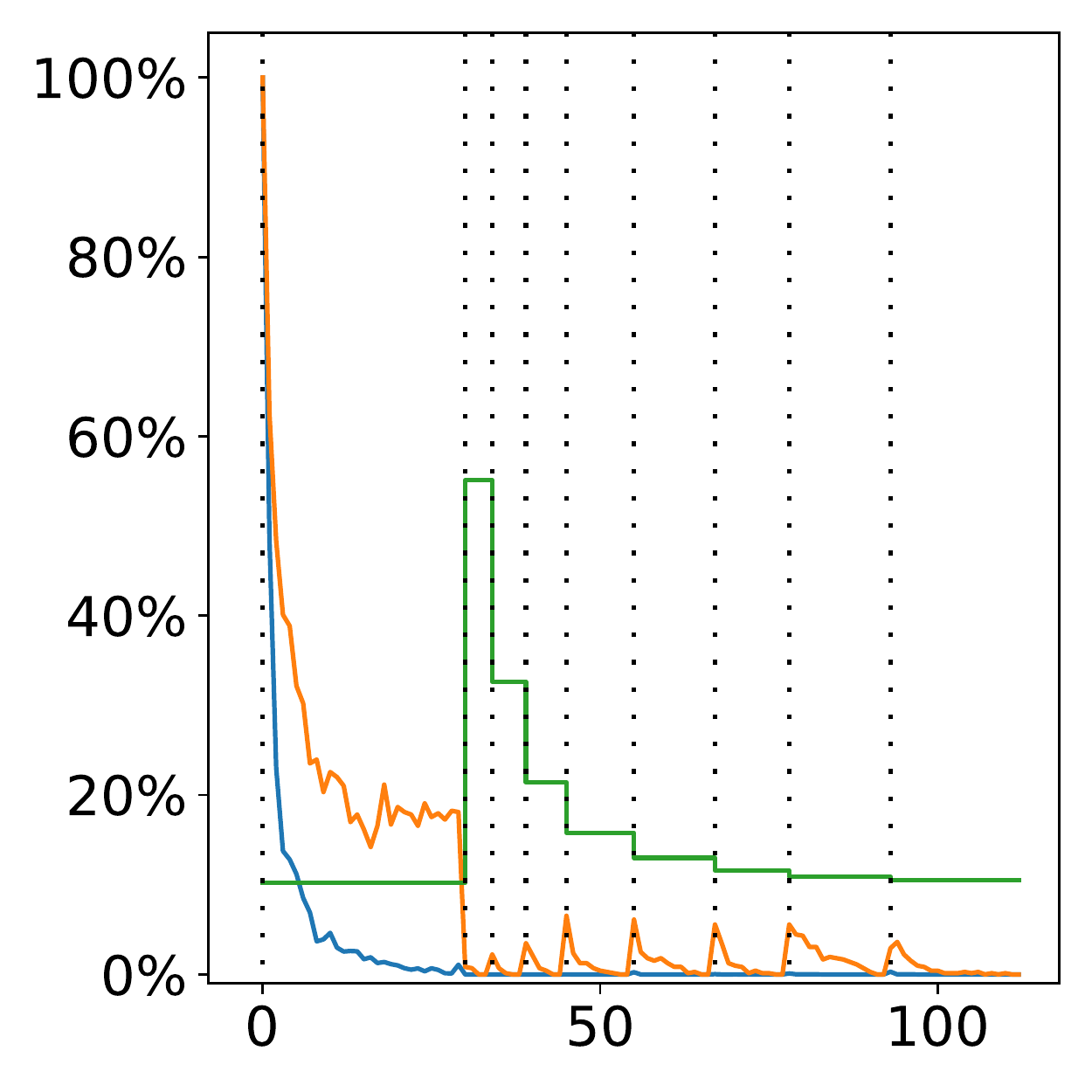} & 
        \includegraphics[width=.31\linewidth]{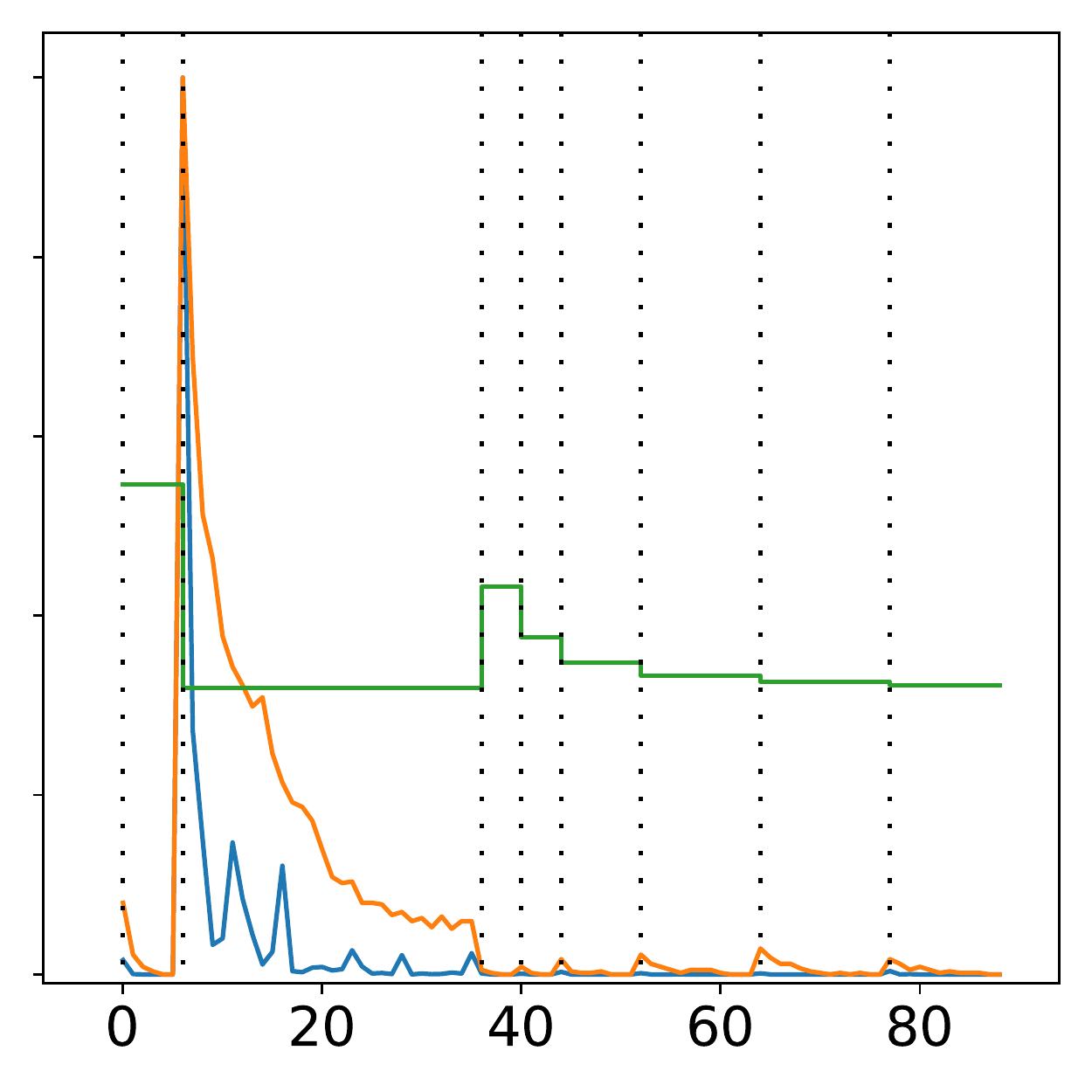} &
        \includegraphics[width=.31\linewidth]{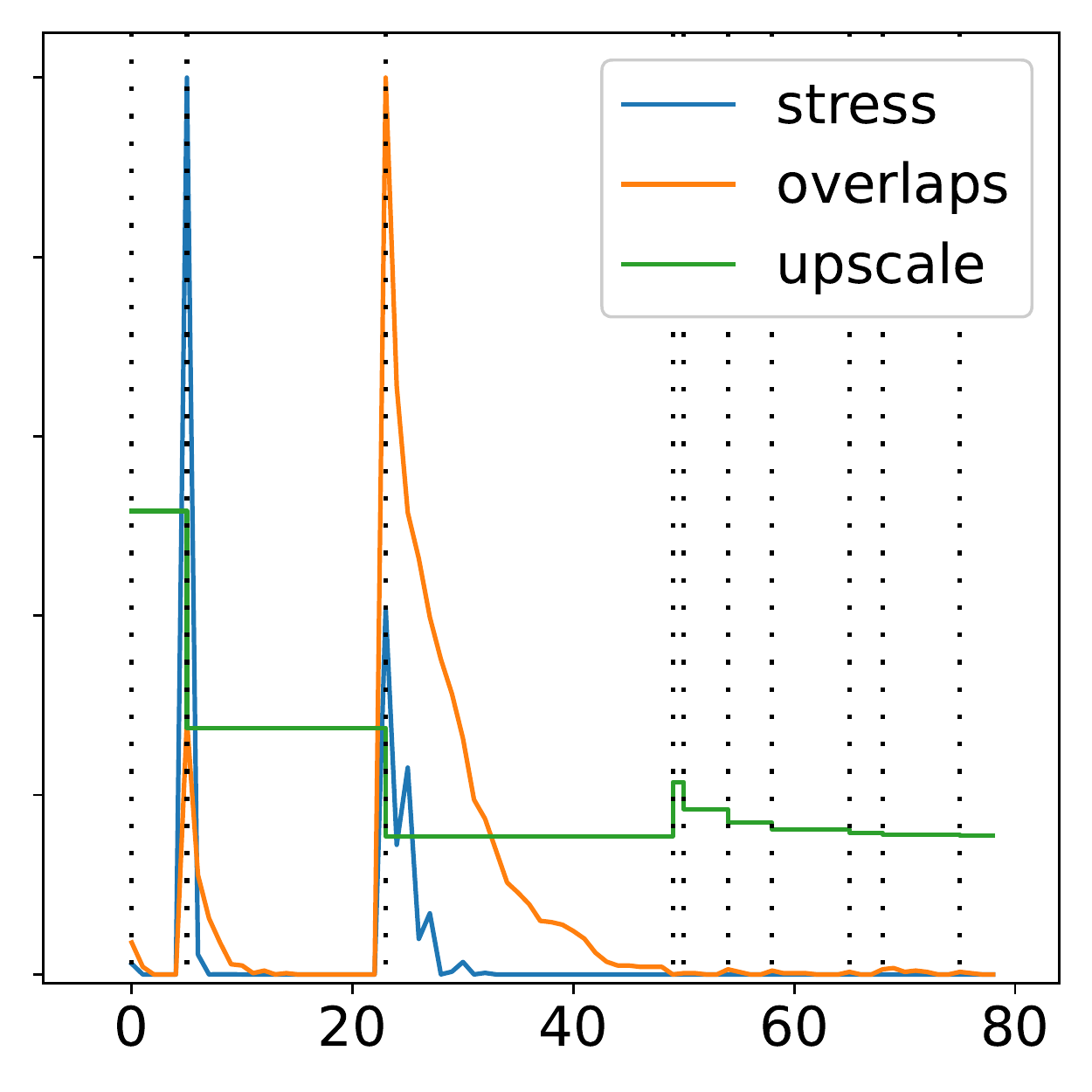} \\
        
        \begin{sideways}{\hspace{1cm}\textbf{\modelnamep}}\end{sideways}&
        \includegraphics[width=.31\linewidth]{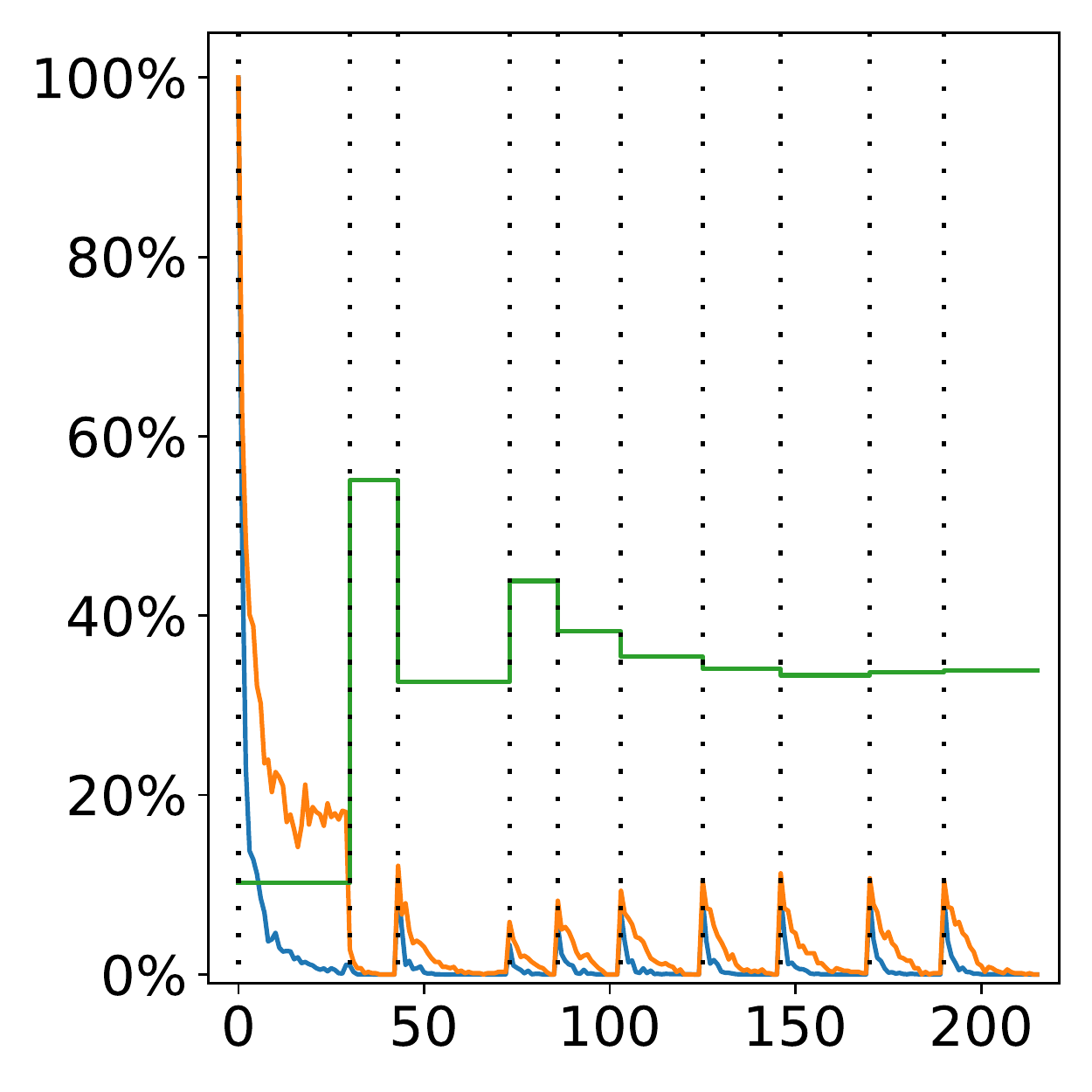} & 
        \includegraphics[width=.31\linewidth]{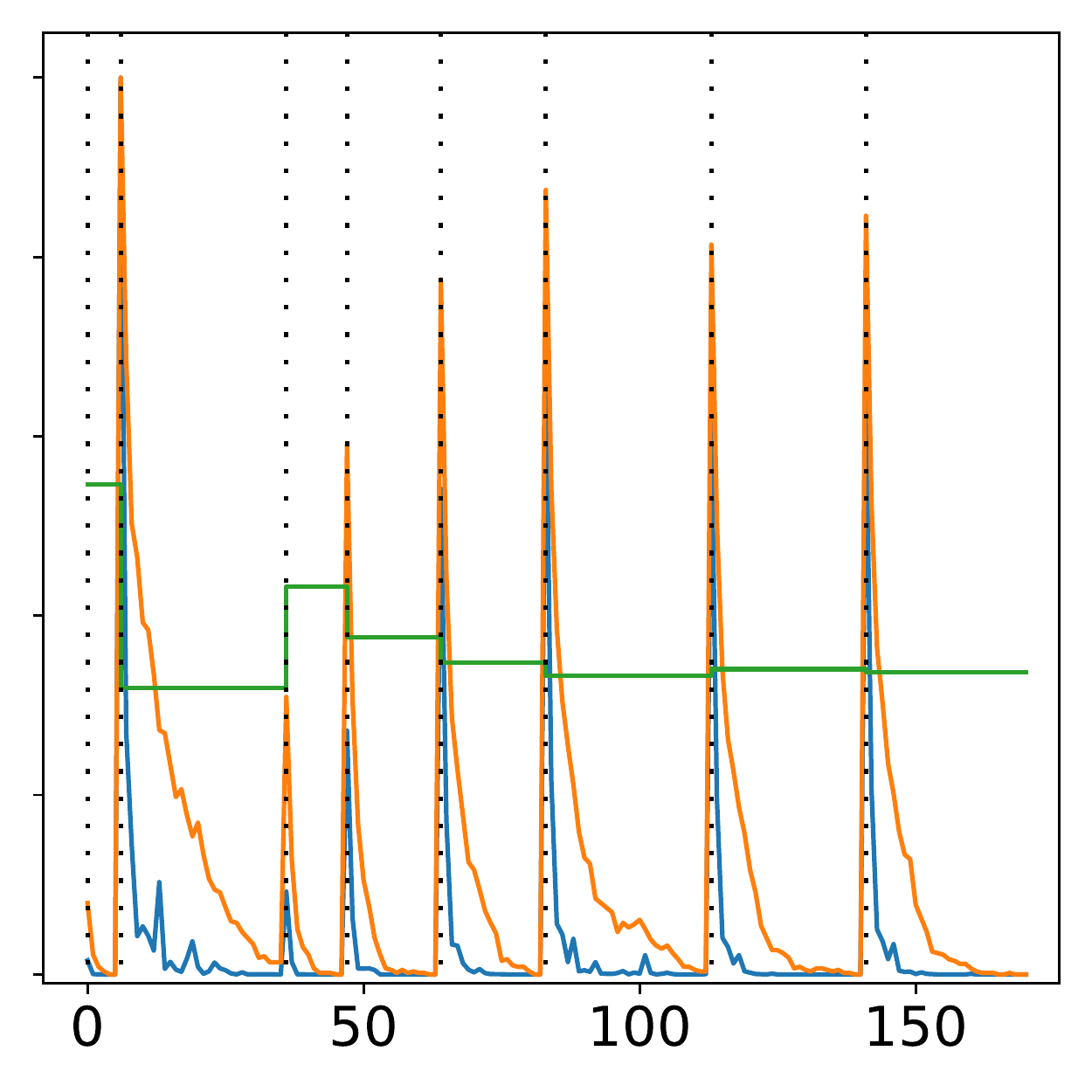} &
        \includegraphics[width=.31\linewidth]{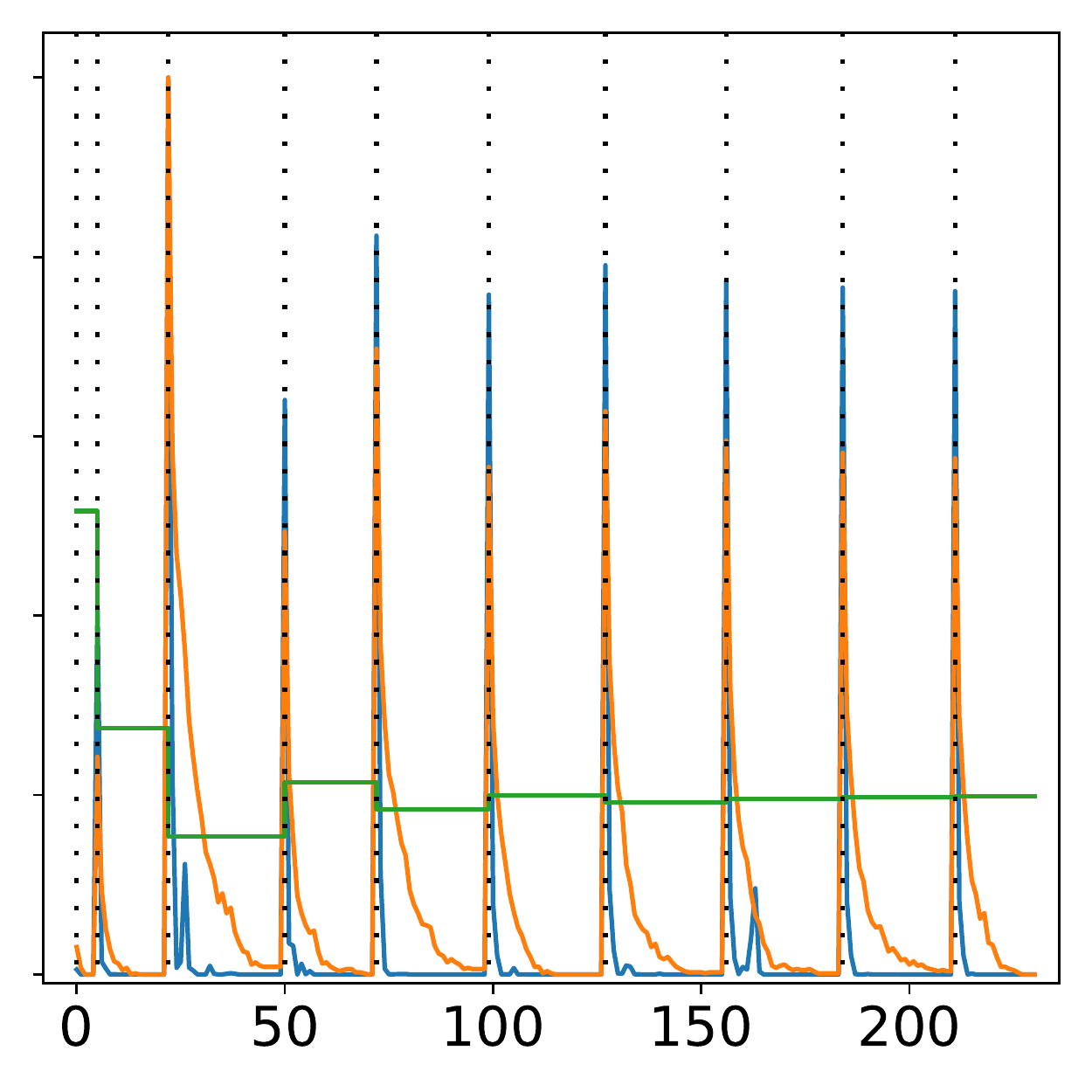} \\
        & & Total number of iterations & \\
    \end{tabular}
    \caption{Convergence plots of \modelname and \modelnamep. It reports the evolution of stress, number of overlaps and scaling ratio against the total number of iterations. Vertical dashed lines represent beginning of new \textit{passes}. Stress and number of overlaps are normalized by their respective maximum value, while upscaling ratio is normalized by its binary search maximum bound (see~\autoref{sec:scaling}).}
    \label{fig:convergence}
\end{figure}

This section studies how re-initializing the layout at every \textit{pass} (\ie call to the optimization algorithm) in \modelnamep affects its convergence speed in comparison to \modelname. Observations of both variants on the \textit{mode}, \textit{badvoro} and \textit{root} \textbf{Graphviz} graphs are presented in \autoref{fig:convergence}. 
Each plot reports the number of passes, the length of each pass and the stress, number of overlaps and scaling ratio evolution against the total number of iterations.

For both \modelname and \modelnamep, the number of passes never exceeded 10. In these executions, the maximum number of iterations in the optimization algorithm was set to 30. Many passes stop before reaching this limit thanks to the stop condition on null nodes movements (see \autoref{sec:sgd2_optim}). 
As the stress and number of overlaps follow the same trends on every plot, it confirms that optimizing our stress effectively removes overlaps. The difference between \modelname and \modelnamep is also distinctly observable. In \modelname, most overlaps are removed in the first few passes while the last ones are dedicated to the search for the optimal upscaling ratio while preserving the overlap-free layout. On the other hand, \modelnamep has to restart from the scaled initial nodes every time a new pass begins, the problem being made simpler or harder through a different upscaling ratio. This explains why \modelnamep is consistently slower than \modelname (see \autoref{sec:exec_time}) but better preserves the initial layout (see \autoref{sec:quantitative_bench}). 

Finally, we would like to discuss the choice of the binary search upper-bound $s_{max}$. In this paper, it was set to the minimum scaling ratio for an overlap-free layout that has not moved any node (see \autoref{sec:scaling}). Doing so guarantees that \modelname finds a overlap-free layout and solve the Overlap Removal task. 
Such an $s_{max}$ value can be high when two nodes are almost perfectly overlapped, but in practice \modelname output layouts scaling remained far from this upper-bound during the benchmark (\eg green lines in \autoref{fig:convergence}). 
Nevertheless, it would be possible to set $s_{max}$ to a lower value to enforce a smaller scaling in the produced layout, at the cost of initial layout preservation and the guarantee that it is overlap-free.
For instance, $s_{max}$ could be set to twice the sum of nodes areas, leading toward a layout which bounding box space is at least half occupied, but there would be no guarantee that this restrained space is enough to provide an overlap-free layout that preserves the initial one. The notion of \textit{overlap} itself could also be approximated with a \textit{tolerance} to speed up the convergence (\ie consider as \textit{not} overlapped the pairs of nodes that do overlap by less than a \textit{tolerance} margin). In the end, these choices mainly depend on the initial graph layout and the desired aspect of overlap-free layout. Although it can be slightly suboptimal, the $s_{max}$ defined and used in this paper (see \autoref{sec:scaling}) is better adapted to the general case.

\section{Conclusion}
\label{sec:conclusion}
This paper has presented \modelname, an Overlap Removal (OR) algorithm that leverages upscaling and stress optimization by simulated stochastic gradient descent to minimize deformations in the initial layout. 
\modelname idea is based on combining upscaling and the preservation of all nodes pairwise distances to produce an overlap-free layout, focusing on the preservation of the initial graph layout structures while limiting the surface used.
It has been compared to several state-of-the-art algorithms, and is among the best techniques to preserve the initial layout, which is critical in graph drawing to retain the readability of the graph layout structures. \modelname complexity is in $\mathcal{O}(s(N^2 +N \log N))$ and is among the fastest methods on the benchmark  graphs (with up to \num{1463} nodes and \num{11582} Overlaps).
Future works leads include improvements of the algorithm complexity to better handle large graphs.
The first idea to achieve that is to sub-sample the nodes pairs to process to remove overlaps. A multi-scale approach could enable to optimize the initial graph layout structures preservation while sampling the distances to preserve (\ie preserve distances between-clusters and within-cluster; ignore between nodes of different clusters).
Finally, with the recent advances in Deep Learning for graph drawing~\cite{giovannangeli2021deep,wang2021deepgd}, we plan to learn a Deep Learning model solve OR problems. By design, these models can scale to large graphs as they are capable of solving the task they have learned in almost constant time.

\section*{Acknoledgements}
We would like to thank the Nouvelle-Aquitaine Region for supporting this work through its foundings OPE 2020-0408 and OPE 2020-0513.

\bibliographystyle{splncs04}
\bibliography{forbid}

\end{document}